\documentclass[superscriptaddress,prx,twocolumn,amsmath,amssymb,nofootinbib]{revtex4-1}
\usepackage[version=4]{mhchem}
\usepackage[T1]{fontenc}
\usepackage{amsthm}
\usepackage{graphicx}
\usepackage{times}
\usepackage{hyperref}
\hypersetup{
colorlinks=true,
linkcolor=red,
citecolor=red,}
\usepackage{textcomp}
\usepackage{enumerate}
\usepackage{multirow}
\usepackage{tabularx}
\usepackage[mathscr]{euscript}
\usepackage{mathtools}
\usepackage{mhchem}
\usepackage[mathscr]{euscript}
\usepackage{dsfont}
\usepackage{amsfonts}
\usepackage[normalem]{ulem}
\usepackage{centernot}
\usepackage{stmaryrd}
\usepackage{longtable}
\usepackage{enumitem,kantlipsum}
\usepackage{array}
\usepackage{wrapfig}
\usepackage{tabu}
\usepackage{graphicx}
\usepackage{placeins}
\usepackage{enumitem}
\usepackage{dutchcal}
\usepackage{tabularx}
\usepackage[export]{adjustbox}
\usepackage{color,soul}
\usepackage{longtable}
\usepackage[dvipsnames]{xcolor}
\usepackage{soul} 

\usepackage[caption=false]{subfig}
\usepackage{adjustbox}

\usepackage{tikz}
\usetikzlibrary{quantikz}

\newcommand\s[1]{_{\rm #1}}

\newcommand{\ketbra}[1]{ | #1 \rangle\!\langle #1 |}

\newcommand{\ii} {\textbf{i}}

\newcommand{\one}{\leavevmode\hbox{\small1\normalsize\kern-.33em1}}

 \def\ee{\mathord{\rm e}}
 
\def\id{\mathord{\rm id}} \def\ii{\mathord{\rm i}}

\renewcommand{\ii}{{\rm i}}
\renewcommand{\ee}{{\rm e}}
\def\beq{\begin{equation}}
\def\eeq{\end{equation}}

\def\barray{\begin{eqnarray}}
\def\earray{\end{eqnarray}}

\usepackage{ulem}

\begin{document}

\title{Suppressing  Amplitude Damping in Trapped Ions: \\ Discrete Weak Measurements for a Non-unitary Probabilistic  Noise Filter}

\author{Andrea Rodriguez-Blanco}
\email{Electronic address: andrer22@ucm.es}
\affiliation{Departamento de F\'{i}sica Teorica, Universidad Complutense, 28040 Madrid, Spain}

\author{K. Birgitta Whaley}
\affiliation{Department of Chemistry, University of California, Berkeley, CA 94720, USA} \affiliation{ Berkeley Center for Quantum Information and Computation, Berkeley, CA 94720, USA}
\author{Alejandro Berm{u}dez}
\affiliation{Instituto de F\'{i}ısica Te\'{o}rica, UAM-CSIC, Universidad Aut\'{o}noma de Madrid, Cantoblanco, 28049 Madrid, Spain}

\date{\today} 

\begin{abstract}
The idea of exploiting maximally-entangled states as a resource lies at the core of several 
modalities of quantum information processing, including secure quantum communication, quantum computation,  and quantum sensing. However, due to imperfections during or after the entangling gates used to prepare such states, the amount of entanglement decreases and their quality as a resource  gets degraded. We introduce a low-overhead protocol to    reverse this degradation by partially filtering out  a specific type of noise relevant to many quantum technologies. We present two  trapped-ion schemes for the implementation of a non-unitary probabilistic  filter against  amplitude damping noise, which can protect any maximally-entangled pair from spontaneous photon scattering during or after the two-qubit trapped-ion entangling gates. This filter  can be understood as  a protocol for single-copy quasi-distillation, as it 
uses only local operations to realise a reversal operation that can be 
understood in terms of weak measurements.
\end{abstract}

\maketitle

\section{INTRODUCTION}
\label{sec:introduction}
Entanglement  allows for new ways of processing and transmitting information in the quantum-mechanical realm, including
quantum teleportation as a paradigmatic example~\cite{PhysRevLett.70.1895}. Given the important role of entanglement in
diverse quantum-information protocols~\cite{nielsen00}, it has become a genuine resource. For instance, maximally-entangled states can be used for secure quantum communications ~\cite{Gisin2007, PhysRevLett.111.100505, Lago-Rivera2021, Ballance_19}, and for quantum sensing and metrology~\cite{Zaiser2016, RevModPhys.89.035002, Pfender2017}, both of which aim at beating the  limitations imposed by the laws of classical physics. In the context of quantum computing, maximally-entangled states between pairs of qubits in a quantum register can be prepared using gates drawn from a universal gate set. Improving the quality of these gates above a so-called fault-tolerance threshold  is crucial to scale up these quantum computers~\cite{Preskill1998, Ladd2010,PhysRevA.52.R2493, nielsen00}.  Unfortunately, the quality of these entangled states, or the gates that produce them, gets degraded by small imperfections in the experimental controls 
as well as by the unavoidable coupling of the system to its surrounding environment. Therefore, a central goal across many quantum technologies is the development of  techniques  to create, store, and distribute maximally-entangled states in the presence of  decoherence and noise.

Amplitude damping is an important mechanism of decoherence 
arising from energy relaxation~\cite{nielsen00} that is common to many  platforms. An example  relevant to the present work is that of trapped-ion optical qubits \cite{PhysRevA.69.032503, PhysRevLett.110.060403, Schindler_2013, PhysRevLett.127.130505}, where spontaneous  emission of photons from a metastable  level leads to  amplitude damping. This 
is summarized by the $T_{1}$-time, which sets the ultimate  decoherence limit for optical qubits
when all other sources of technical noise are suppressed. For trapped-ion hyperfine or Zeeman qubits \cite{ PhysRevLett.117.060505, PhysRevLett.117.060504}, where the information is encoded in the groundstate manifold, spontaneous photon scattering during storage vanishes. However, such photon scattering becomes relevant when creating and manipulating the entangled states, e.g., when  
using two-photon Raman transitions via auxiliary excited states, during which a residual emission of photons can contribute unfavourably to the gate fidelities~\cite{Wineland1998,  PhysRevA.75.042329}. Spontaneous emission can also be a limitation in Rydberg-atom quantum processors, where in order to achieve high two-qubit gate fidelities, long coherent ground-Rydberg state Rabi oscillations are needed. However, the presence of spontaneous emission decay channels from the intermediate excitation state to the ground manifold, and from the target Rydberg state to lower-$n$ Rydberg states, with $n$ the principal quantum number, limit the coherence and population times \cite{PhysRevA.97.053803,PhysRevA.101.062309,Bluvstein2022}.\\
\indent
To fight against  amplitude damping or, indeed any  source of decoherence, one may redundantly encode the quantum-information into logical qubit by using more physical qubits. The theory of quantum error correction (QEC) shows that it is possible to  exploit  multi-partite entanglement among the encoded physical qubits to actively detect and correct the errors that have occurred on the logical qubits without actually perturbing the encoded information~\cite{nielsen00, doi.org/10.48550/arxiv.quant-ph/0004072,  lidar_brun_2013, doi:10.1080/00107514.2019.1667078}.
In recent years, we have witnessed a remarkable progress in experimental QEC, especially in  trapped-ion and superconducting-circuit platforms. We have seen how one logical qubit can be protected from an  arbitrary error using 7 physical qubits for the color QEC code~\cite{doi:10.1126/science.1253742, PhysRevX.11.041058,Postler2022}, or 9 \cite{PhysRevLett.129.030501, Krinner2022}, and 25 physical qubits~\cite{https://doi.org/10.48550/arxiv.2207.06431} for the surface QEC code. Moreover, the advantage of using  fault-tolerant designs has also been demonstrated in~\cite{Postler2022,https://doi.org/10.48550/arxiv.2208.01863}. This has allowed, for the first time, to realize a full universal gate set at the logical level, including transversal gates that create a logical entangled state using 14 physical qubits~\cite{Postler2022,https://doi.org/10.48550/arxiv.2208.01863}.  
QEC strategies are known to provide a scalable  solution to build large fault-tolerant quantum computers~\cite{doi:10.1137/S0097539799359385, RevModPhys.87.307,royalsocietypublishingResilientQuantum} that starts to become practically relevant as the experimental technologies increase the possible qubit redundancy.
However, to achieve a significant advantage of quantum encoding when the noise and control errors lie in the vicinity of the aforementioned fault-tolerance threshold, even the most promising schemes of quantum error correction (QEC)~\cite{lidar_brun_2013, doi:10.1080/00107514.2019.1667078, doi.org/10.48550/arxiv.quant-ph/0507174}  typically require a very high degree of  redundancy 
which leads to  large  overheads in the number of physical qubits~\cite{RevModPhys.87.307}.
Thus there is still a long road ahead for truly large-scale QEC. In the meantime, it is important to develop alternative schemes that reduce the effect of noise with a lower qubit overhead. Moreover, some of these noise mitigation techniques  could  be eventually combined with QEC.

Some of these alternative strategies  work best for specific types of noise. For instance,  dynamical decoupling~\cite{PhysRevA.58.2733} refocuses the effects of dephasing noise caused by external fields with sufficiently-slow fluctuations, whereas decoherence-free subspaces~\cite{PhysRevLett.81.2594} exploit symmetric subspaces that are immune to external fields with sufficiently-global fluctuations. In the context of QEC, the qubit overhead can also be reduced if one focuses on a  set of errors that is believed to be the main noise source in a specific platform. 
Thus, one may devise channel-adapted QEC codes for amplitude damping by using four qubits to correct for one error~\cite{PhysRevA.56.2567,4675715}. In this work, we explore a different error suppression strategy that is framed in the context of  {\it entanglement distillation}~\cite{D_r_2007}, which includes schemes for entanglement concentration and entanglement purification as specific limits, and will allow us to further minimize the qubit overhead. Such schemes are particularly relevant when the entanglement is distributed between a pair of spatially-separated physical qubits. 

For bipartite systems under realistic gates and channels,  the prepared maximally-entangled pairs are 
neither perfectly transformed under quantum operations nor perfectly distributed to distant parties. 
Entanglement distillation aims at exploiting local operations and classical communication (LOCC) to improve the fidelity of a collection of noisy partially-entangled  mixed states with respect to a target maximally-entangled pure state~\cite{PhysRevLett.76.722}. We note that this question is not only of practical relevance in some applications, e.g., quantum repeaters for  quantum communications over large distances~\cite{PhysRevLett.81.5932}, but has also played a key role in the development of the current understanding  of entanglement via resource theories. Specifically, in a theoretical framework one considers extracting $m(n)$ maximally-entangled pairs $\ket{\Psi_{\rm AB}}$ from  $n$ copies of a partially-entangled mixed state $\rho_{\rm AB}$, by using a suitable  LOCC protocol $\mathcal{E}\in{\rm LOCC}$. The efficiency of the distillation protocol is specified by the asymptotic ratio $m(n)/n\to\xi_{\rm D}$ in the limit of an infinite number of copies $n\to\infty$. This allows to define the so-called distillable entanglement $E_{\rm D}(\rho_{\rm AB})={\rm sup}_{\mathcal{E}\in{\rm LOCC}}\{\xi_{\mathit{ D}}\}$,  which corresponds to the optimal efficiency for all conceivable distillation protocols. This yields an operational measure of entanglement~\cite{10.5555/2011706.2011707,Aolita_2015}, quantifying the amount of entanglement in a  state by the efficacy with which one can perform a particular task, namely to distill perfect maximally-entangled pairs from it.

From this theoretical perspective, distillation protocols aim at producing maximally-entangled pairs with unit fidelity in such an asymptotic limit. In a seminal work~\cite{PhysRevLett.76.722}, Bennett {\it et al.} showed that  collective measurements on the qubits  belonging to each of the parties separately, allow for a LOCC scheme that can distill a non-vanishing number of maximally-entangled pairs from  $n\to\infty$ copies, provided that the initial fidelity of the noisy mixed state is above $1/2$. It was then shown  that this constraint on the initial state can be lifted, provided that one has previously applied a filtering operation to each of the separate copies~\cite{PhysRevLett.78.574}. This raised the question of exploring the capabilities of single-copy distillation schemes, which would thus reduce  the large overhead in   the number of partially-entangled copies. To our knowledge, the first such scheme finds its root in the work on entanglement concentration for pure states by filtering
~\cite{PhysRevA.53.2046}, the so-called \textit{Procrustean} method, which also applies to certain partially-entangled mixed states~\cite{GISIN1996151}. We note that this single-copy distillation scheme has been realised in photonic experiments~\cite{Kwiat2001}, where one aims at achieving the highest-possible fidelities  by post-selecting on the experimental outcomes of a generalised measurement~\cite{NIELSEN2002249}. In this context,  the aforementioned theoretical limit of a perfect distilled state is unreachable, as the post-selection probability  drops to zero. In any case, since the filtering  operations are always noisy in a practical experiment, the limit of perfect distillation is an idealization even for multi-copy distillation schemes.

The link between quantum error suppression and entanglement distillation derives from the fact that single-copy distillation methods fall in the class of probabilistic quantum error detection (pQED) \cite{PhysRevLett.82.2598}. After the measurement, one can infer whether an  error, e.g., amplitude-damping, has occurred  or not, and then keep those outcomes where the encoded information can be probabilistically recovered by simply reversing the effect of the measurement~\cite{PhysRevLett.68.3424,PhysRevLett.82.2598}.  
The first  proof-of-principle demonstrations of such pQED was performed in photonic systems using projective measurements implemented via photon absorption~\cite{Kwiat2001, Pan2003, Yamamoto2003, PhysRevLett.90.207901}. As remarked in~\cite{Reichle2006}, this absorptive measurement  destroys the quantum state, such that the distilled entanglement cannot be used for any subsequent quantum-information task. In contrast, the trapped-ion experiment of~\cite{Reichle2006} allowed demonstration of a two-copy distillation protocol that improves the quality of a single entangled pair which is not destroyed and thus remains available for further posterior processing. In a subsequent trapped-ion experiment~\cite{PhysRevLett.111.180501}, a shelving mechanism using additional states of the trapped-ion level structure was exploited to  detect leakage without disturbing the qubit computational states. In the current work, we are interested in designing single-copy distillation schemes to fight against amplitude damping and protect any maximally-entangled Bell pair  created using trapped-ion technologies.

Although  generic mixed states are not useful for  perfect distillation~\cite{PhysRevLett.81.2839,PhysRevLett.81.3279},  there are certain families of them that can be distilled arbitrarily close to the limit of  unit fidelities~\cite{PhysRevA.60.1888}. This so-called {\it quasi-distillation} is closer in spirit to the experimental situation, where the filtering operations are never perfect.  As shown in~\cite{PhysRevA.64.010101}, one can find optimal single-copy quasi-distillation  protocols with  specific filtering operations that correspond to generalised measurements, and depend on the  form of the initial  mixed state. These optimal strategies thus require prior information about the initial  state, a property that is shared with the original Procrustean methods~\cite{PhysRevA.53.2046, GISIN1996151}, and subsequent works ~\cite{Kwiat2001, Ota_2012, Liao_2013}. In this work, we are interested in  single-copy quasi-distillation schemes where the filtering  requires a priori information that is independent of the maximally-entangled state one wants to distill. 
We show that this is possible for specific noise channels, where the prior information now depends on the  noise.  We present a low-resource probabilistic method to protect an unknown entangled pair against amplitude damping, which acts as a noise filter and is related to the Procrustean method of entanglement concentration~\cite{PhysRevA.53.2046}. The scheme presented here exploits a specific form of measurement reversal with origins in the context of weak measurements~\cite{PhysRevLett.97.166805, PhysRevA.82.052323}.  In addition to the focus on explicit mitigation of errors due to amplitude damping rather than on entanglement concentration in general, a technical difference from ~\cite{PhysRevA.53.2046} is that the present scheme does not require any prior knowledge of the target state, which can be any of the maximally-entangled states, but instead requires as an input parameter the  $T_{1}$ time associated with the amplitude-damping noise. This can be determined from previous calibration experiments \cite{PhysRevLett.110.060403, PhysRevA.69.032503}, and then fed into the probability of amplitude damping $p$ for each qubit~\cite{Preskillnotes}, namely
\beq
\label{eq:p_T1}
p=1-\ee^{- \frac{t}{T_{1}}}.
\eeq

We provide two possible  schemes for a trapped-ion implementation of this quasi-distillation protocol. The first one is  related to the idea of quantum logic spectroscopy~\cite{Schmidt749}, and exploits unitaries between the qubits and some of the common vibrational modes of the ion crystal holding them. The second method, on the other hand, exploits  phonon-mediated entangling gates~\cite{HOME2013231, Bruzewicz2019,  PhysRevA.105.022623} to map the relevant information from the physical qubits onto the ancillas. The latter approach turns out to be more robust with respect to thermal fluctuations in the common vibrational modes. As noted above, this method can protect any unknown entangled pair, or even be applied at the level of the full entangling unitary that prepares such entangled pairs. Both methods exploit   a   measurement in which the  physical qubit effectively interacts with an ancilla qubit,  such that the ancilla gets flipped when an amplitude-damping error occurs, and can be used to design an operation that reverses the amplitude damping. The probabilistic character 
 arises from the fact that we keep (post-select) those states for which the ancilla measurement  signals "no error". The method can thus also be viewed as acting as an amplitude-damping noise filter.

The remainder of the article is organized as follows. In Sec. \ref{sec:general_scheme}, we present the general scheme to show how the filtering protocol based on quantum measurement reversals  for amplitude damping can be used to distill single-copy entangled states. In Sec. \ref{sec:trapped_ions} we then propose two different schemes for the experimental implementation of the quantum measurement reversals in trapped-ion platforms. Adapted to the trapped-ion native logic operations, the first method in Sec.~\ref{sec:schemeA} relies on quantum logic spectroscopy techniques mediated by common vibrational modes. The second method in Sec.~\ref{sec:schemeB} consists on a sequence of one and two-qubit gates, both of which are quite robust to the thermal motion of the ions. Finally, in Sec.~\ref{sec:results}, we benchmark both schemes while highlighting the power to protect any entangled state by calculating  the average gate fidelity, and discuss their practical limitations under more realistic conditions. 
Sec.~\ref{sec:conclusion} summarizes and presents an outlook for further development.

\section{Noise filtering against amplitude damping}

\subsection{General scheme}\label{sec:general_scheme}

\begin{figure}[ht!]
    \includegraphics[width=1\columnwidth]{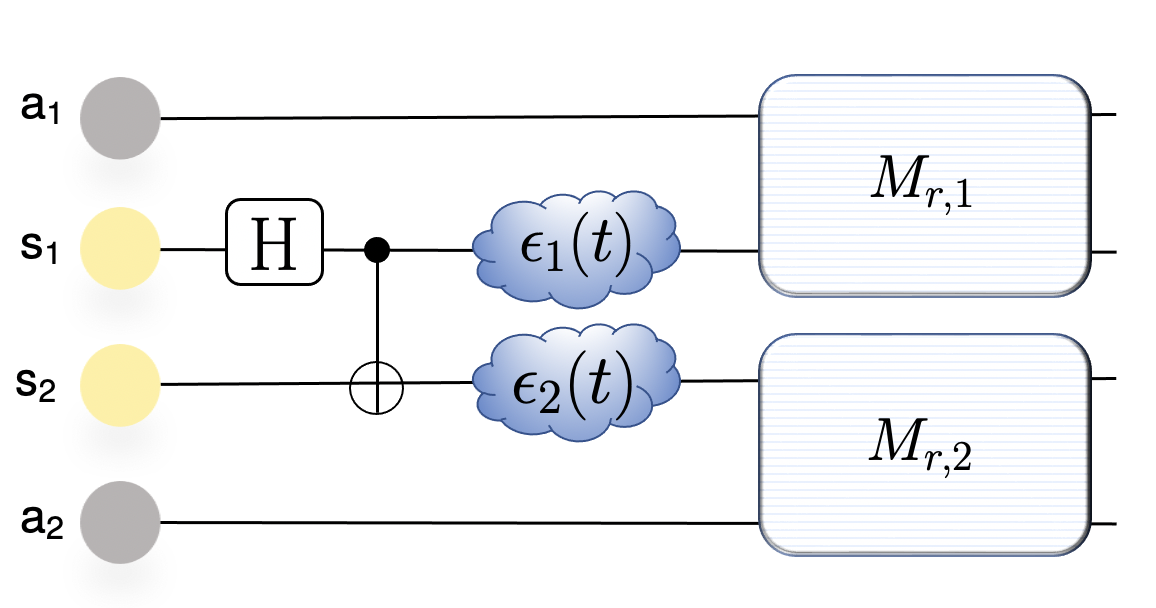}\\
    \caption{{\bf Amplitude-damping reversal for single-copy distillation.} We consider a pair of physical or system qubits ${ s}_1,{ s}_2$, which can be prepared in a maximally-entangled Bell pair~\eqref{eq:bell_pairs}  via a Hadamard gate $H$ and an entangling CNOT gate, as shown in the circuit. The system qubits are then subjected to uncorrelated amplitude damping channels, Eq.~\eqref{eq:damping_composition}, which may model the main error source during the gate or the primary source of environmental noise during a waiting period $t$. The resulting partially-entangled mixed state can be distilled into a state with a higher fidelity with respect to the targeted Bell pair via the reversal operations $M_{r,1},M_{r,2}$, Eq.~\eqref{eq:reversal_operations}, which act as non-unitary filters and must be implemented by coupling the qubits to ancillary degrees of freedom $a_1,a_2$, and post-selecting on specific outcomes of subsequent measurements.  }.
    \label{fig:general_filtering_scheme} 
\end{figure}

Consider a pair of qubits ${\rm s}_1,{\rm s}_2$ that can be prepared in any  Bell pair by a maximally-entangling unitary $U_{\rm id}$. Using the first two unitary gates of the circuit displayed in Fig.~\ref{fig:general_filtering_scheme} with $U_{\rm id}=U_{\rm CNOT}$, we have
\beq
\begin{split}
U_{\rm id}\ket{\pm}_{{\rm s}_1}\!\!\otimes\ket{0}_{{\rm s}_2}=\ket{\Phi_{\pm}}=\textstyle{\frac{1}{\sqrt{2}}}\left(\ket{0}_{{\rm s}_1}\!\!\otimes\ket{0}_{s_2}\pm\ket{1}_{s_1}\!\!\otimes\ket{1}_{{\rm s}_2}\right),\\
U_{\rm id}\ket{\pm}_{{\rm s}_1}\!\!\otimes\ket{1}_{{\rm s}_2}=\ket{\Psi_{\pm}}=\textstyle{\frac{1}{\sqrt{2}}}\left(\ket{0}_{{\rm s}_1}\!\!\otimes\ket{1}_{{\rm s}_2}\pm\ket{1}_{{\rm s}_1}\!\!\otimes\ket{0}_{{\rm s}_2}\right).
\end{split}
\label{eq:bell_pairs}
\eeq
As outlined in the introduction, we aim at designing a probabilistic filtering method that, in contrast to previous entanglement concentration schemes \cite{PhysRevLett.76.722, Reichle2006} can protect any of these maximally-entangled states and requires  prior information about the noise instead of the particular state. We note that this entangling operation might be any other unitary gate  native to the specific experimental setup. In fact, the scheme can also be applied to quantum network scenarios  where entanglement between distant qubits is heralded via photonic interconnects, as has been demonstrated for trapped ions~\cite{Moehring2007,PhysRevLett.100.150404,Pironio2010, Hucul2015,PhysRevLett.124.110501,Nadlinger2022,nichol2021quantum}. In that case, the scheme below serves to quasi-distill a specific target entangled pair from the heralded two-qubit state.

Our scheme thus serves to protect the full entangling unitary $\rho_{\rm id}=U_{\rm id}\rho_0U_{\rm id}^\dagger$ against  an amplitude damping channel
\beq
\label{eq:damping_composition}
\rho_{\rm id}\mapsto\tilde{\rho}_{\rm id}=\epsilon(\rho_{\rm id})=\epsilon_{{\rm s}_1}\circ\epsilon_{{\rm s}_2}(\rho_{\rm id}),
\eeq
which is represented by the shaded clouds in Fig.~\ref{fig:general_filtering_scheme}. Here
 we have defined
\begin{equation}
\label{eq:AD_channel}
\epsilon \s{s_{q}}(\rho_{\rm id})=K\s{0, s_{q}}^{\phantom{\dagger}}\rho_{\rm id}K\s{0, s_{q}}^\dagger+K\s{1, s_{q}}^{\phantom{\dagger}}\rho_{\rm id}K\s{1, s_{q}}^\dagger,
\end{equation}
in terms of the Kraus operators $K_{0,{\rm s}_1}=K_0\otimes\mathbb{I}_2,K_{0,{\rm s}_2}=\mathbb{I}_2\otimes K_0$,  $K_{1,{\rm s}_1}=K_1\otimes\mathbb{I}_2,K_{1,{\rm s}_2}=\mathbb{I}_2\otimes K_1$, with
\begin{equation}
    K_{0} = 
    \begin{pmatrix}
    1 & 0 \\
    0 & \sqrt{\bar{p} }
    \end{pmatrix},\hspace{0.1cm} K_{1} = 
    \begin{pmatrix}
    0 & \sqrt{p} \\
    0 & 0 
    \end{pmatrix},
    \label{eq:Krauss}
\end{equation}
where $p$ is the probability of an amplitude damping error and $\bar{p}=1-p$ is the probability that no amplitude-damping error occurs, with $p$ dependent on the $T_1$ time via Eq.~\eqref{eq:p_T1}. These single-qubit Kraus operators defined in the computational basis represent the amplitude damping noise channel \cite{nielsen00,4675715}. This is an asymmetric channel since the qubit states $\ket{1}\s{s_q}$ are transformed to $\ket{0}\s{s_q}$ with probability $0<p\leq 1$, whilst the $\ket{0}\s{s_q}$ states never transform into $\ket{1}\s{s_q}$, regardless of the value of $p$. As noted above, the amplitude-damping probability can be related to the natural lifetime of the qubit, due e.g., to photon scattering from a metastable level in trapped-ion optical qubits, or to the residual photon-scattering when the entangling gate is mediated by two-photon processes via far-detuned dipole-allowed transitions.

If the decoherence due to  amplitude damping is  caused by photon emission into the  electromagnetic (EM) environment, we can effectively understand the decay channel as a partial (often referred to as a weak) measurement of the photon number exerted by the aforementioned environment \cite{Preskillnotes}. The partial collapse of the amplitudes from the excited to the ground state can then be seen as the result of weak measurements performed by the independent EM modes that can absorb the emitted photon by each of the system qubits. 
In contrast to the more common von Neumann projections \cite{nielsen00}, a weak measurement does not fully collapse the quantum state into an eigenstate of the operator being measured, and is thus reversible \cite{PhysRevLett.82.2598,KOFMAN201243,Kim:09}. Therefore, for decay probabilities $0<p<1$, a secondary weak measurement can be applied to reverse the partial collapse, leading to a quantum measurement reversal of the aforementioned errors \cite{Kim:09}.\\
\indent
This reversal can be accomplished by a filtering operation. In the case of amplitude damping, this corresponds to the  last pair of operations depicted in Fig.~\ref{fig:general_filtering_scheme}, which read
\begin{eqnarray}
\tilde{\rho}_{\rm id}\mapsto M_{r,2}^{\phantom{\dagger}}M_{r,1}^{\phantom{\dagger}}\tilde{\rho}_{\rm id} M_{r,1}^{\phantom{\dagger}}M_{r,2}^{\phantom{\dagger}},
\nonumber \\
M_{r, q}=
\frac{1}{\sqrt{P_{r_q}}}\begin{pmatrix}
\sqrt{\bar{p}_{r_{q}}} & 0 \\
0 & 1
\end{pmatrix}.
\label{eq:reversal_operations}
\end{eqnarray}
 Note that these reversal operations are applied to each of the physical qubits forming the partially-entangled pair. As discussed in detail below, to implement such a non-unitary filter,  the physical qubits must be coupled to an ancillary subsystems, which must then be measured.  In Eq.~(\ref{eq:reversal_operations}), $P_{r_q}$ is the post-selection success probability.
We note that the fidelity of the recovery process will never be strictly equal to one, as it is a non-trace-preserving map \cite{ PhysRevLett.82.2598, PhysRevLett.97.166805}.  

Let us illustrate how the filtering method works for a particular entangled state. Consider the circuit shown in Fig. \ref{fig:general_filtering_scheme} with the system qubits initialized in the tensor product state $\ket{1}_{{\rm s}_1}\!\!\otimes\ket{1}_{{\rm s}_2}$ such that the Hadamard gate on the first qubit and the unitary $U_{\rm id}=U_{\rm CNOT}$  create the pure maximally-entangled state $\rho_{\rm id}=U_{\rm id}\rho_0U_{\rm id}^\dagger=\ket{\Psi_{-}}\bra{\Psi_{-}}$. The system qubits can undergo amplitude damping either during the gate, or after a subsequent memory time $t$.
Assuming that both qubits are identical and subjected to the same uncorrelated noise environments, the decay probabilities can be considered to be the same, $p_{1}=p_{2}=p$, and the density matrix evolves to the partially mixed entangled-state
\beq
\rho_{\rm id}\mapsto\tilde{\rho}_{\rm id}=p\ket{0,0}\bra{0,0}+\bar{p}\rho_{\rm id}.
\eeq
Recalling that $\bar{p}=1-p$, we see that with probability $p$ the amplitudes of the excited states on the entangled pair decay to the two-qubit tensor product ground state $\ket{0,0}=\ket{0}_{{\rm s}_1}\!\!\otimes\ket{0}_{s_2}$, while with probability $1-p$, the  system remains in the pure maximally entangled state. The bare/unfiltered state infidelity in this situation is 
equal to $p$, i.e.,
\beq
\label{eq:infidelity_unfiltered}
\varepsilon_{\rm unf}=1-\mathcal{F}_{\rm unf}= 1-\bra{\Psi_-} \tilde{\rho}_{\rm id}\ket{\Psi_-}=p.
\eeq

We now use the amplitude damping filtering protocol to distill a single-copy and increase the fidelity with the target entangled state. As emphasized above, our scheme is independent of the initial maximally-entangled state  and it is also valid for any entangling gate, in particular for the Mølmer-Sørensen (MS) gates of trapped ion architectures that we will use later on. The reason for this generality is that the protocol focuses on removing the additional amplitude that decayed from the excited states to the ground states. The amount that needs to be removed is proportional to the decay probability of Eq.~(\ref{eq:p_T1}) and thus it is only the specific value of $T_{1}$ that is  required as prior  information  to implement the single-copy distillation. The decay removal procedure is carried out by applying a quantum measurement reversal on each qubit, with equal strengths $p_{r_{1}}=p_{r_{2}}=p_{r}$, such that the density matrix transforms according to Eq.~\eqref{eq:reversal_operations}. This leads to
\beq
\label{eq:density_filtered_1}
\begin{split}
\tilde{\rho}_{\rm id}\mapsto \rho_{\rm f}
 =\frac{1}{P_{r}}(\bar{p}_{r}^{2}p\ket{00}\bra{00}+\bar{p}_{r}\bar{p}\rho_{\rm id}), 
\end{split}
\eeq

where  $P_{r}=\bar{p}_{r}^{2}p+\bar{p}_{r}\bar{p}$ is a normalization constant that represents the success probability of the measurement reversal. The probability of decaying to the common ground state is now $\bar{p}_{r}^{2}p/P_{r}$, and the probability of remaining in the maximally-entangled state is now $\bar{p}_{r}p/P_{r}$. By setting the strength of the reversal operation equal to the probability of amplitude decay (Eq. (\ref{eq:p_T1})), namely $p_{r}=p$, we get
\beq
\label{eq:density_filtered}
\begin{split}
 \rho_{\rm f}=\frac{1}{\bar{p}^{2}(p+1)}\big(\bar{p}^{2}p\ket{00}\bra{00}+\bar{p}^{2}\rho_{\rm id}\big).
\end{split}
\eeq
We see that the probability of decaying to the ground state has now been effectively reduced by a factor of $1/(1+p)$. This increases the filtered fidelity by the same factor, so that 
\beq
\label{eq:infidelity_filtered}
\varepsilon_{\rm f}=1-\mathcal{F}_{\rm f}= \frac{p}{1+p}.
\eeq

This example illustrates  how the filtering protocol can be used for single-copy  quasi-distillation. The distillation power increases with increasing $p$, i.e., with increasing time $t$, since the infidelity ratio $\varepsilon_{\rm f}/\varepsilon_{\rm unf}=1/(1+p)$ decreases with $p$. However, the probability of successful  filtering, $P_{r}$,  also decays with time according to its dependence on $p$ (see below Eq.~(\ref{eq:density_filtered})). For instance, after two different combinations of gate/memory times, such as $t/T\s{1}=1/10$ and $t/T\s{1}=1$ with associated decay probabilities of  $p(1/10)=0.01$ and $p(1)=0.63$, the corresponding probabilities of quasi-distillation success will be $P_{r}\approx 0.90$ and $P_{r}\approx0.20$, respectively. This clearly shows  how the probabilistic nature of the method arises, as well as the trade-off between a higher quality distillation and a more frequent distillation. For a number  of repetitions 
 $N= 100$, we will be filtering out part of the amplitude damping correctly in 90 instances for the first case, whilst for the second case, the correct instances reduce to  20. Consequently, for a given number $N$ of experimental runs, the greater the amplitude decay probability $p$, the fewer the successful events where the noise is suppressed. \\

Having presented and illustrated the scheme in this section, the remaining task is to describe how one can implement such filtering operations in practical setups. In the context of weak measurements, employing a quantum measurement reversal to reverse the effect of noise can be realized by first applying a weak measurement and then applying the reversal operation.
This idea has been previously addressed both theoretically and experimentally for single qubits, in solid-state systems \cite{PhysRevLett.97.166805}, superconducting qubits \cite{Katz1498,PhysRevLett.101.200401,Zhong2014},  trapped ions~\cite{PhysRevLett.111.180501,PhysRevA.87.012131} and photonic systems \cite{Kim2012,Lee:11}.
However, the reversal of a partial collapse on entangled states has only been  considered for  
photonic systems~\cite{Kim2012,PhysRevA.86.032304,PhysRevA.86.012325,Wang2016,PhysRevA.82.052323}. 
In the next section we develop a filtering scheme based on weak quantum measurement reversal for single-copy distillation of any entangled state for trapped-ion platforms. The low qubit overhead of our scheme makes it interesting as an alternative to standard QEC, since it  can effectively suppress the effect of noise on platforms and/or applications which at present are only capable of manipulating a relatively small number of qubits. Moreover, it is also  practical for  communication situations, where the physical qubits are held by distant parties, and conventional QEC is not straightforward.

The full protocol  to implement this non-unitary filter  is based on the construction of asymmetric Positive Operator-Valued Measure (POVM) operators followed by postselection on ancilla qubits. Section~\ref{sec:trapped_ions} introduces two different schemes to perform the necessary asymmetric POVMs. 
The first scheme builds on the work presented in~\cite{PhysRevA.87.012131}, where a theoretical formalism to implement  symmetric POVMs for trapped ions using quantum logic spectroscopy (QLS) operations~\cite{Schmidt749} was discussed. Here, we extend this QLS scheme  to the design of asymmetric POVM operators, which will be crucial to exploit them as non-unitary filters for single-copy distillation against amplitude damping. The main difference between symmetric and  asymmetric POVMs is that, whilst in the symmetric case one always recovers the projectors onto computational basis  for the extremal cases with $p_{r}=0$ and $ p_{r}=1$, in the asymmetric case one recovers either the identity r a single projection operator in these limits, but never two projectors. 

As discussed in detail in Sec.~\ref{sec:schemeA}, this QLS scheme exploits  common vibrational modes between the system and ancillary qubits, which should be previously laser cooled to the vibrational ground state. This makes non-unitary filtering susceptible to thermal fluctuations in the common vibrational mode, compromising in this way the efficacy of the distillation method. We present a detailed account of this error source below. In Sec.~\ref{sec:schemeB} we then present the second scheme that is not ultimately restricted to the ions operating in the vacuum vibrational mode. Similarly inspired by the case of symmetric POVMs~\cite{PhysRevA.87.012131}, we show that now the asymmetric reversal of amplitude damping can be implemented in terms of  single and two-qubit gates~\cite{Leibfried2003}, which are far more robust to the specific  motional state of the ions. As discussed in detail below, these gates must be followed by projective measurement and post-selection on the ancilla qubits. In Sec.~\ref{sec:results} we compile the resulting circuits into native trapped-ion gates and compare the average gate fidelities of a two-qubit gate in the presence and absence of the filtering process.

\section{Implementation of quantum measurement reversal for trapped ions}
\label{sec:trapped_ions}
In this section, we  give a detailed account of the two schemes for implementation of asymmetric POVMs in trapped ion architectures, and discuss how these connect to the non-unitary filters of Eq.~\eqref{eq:reversal_operations} for single-copy quasi-distillation. Common to both schemes  is the need to use one ancilla qubit per system qubit in order to perform the  probabilistic error detection. Thus, for the two-qubit maximally entangled state, two ancilla qubits must be added to the qubit register. We now discuss how the common vibrational modes, togther with projective measurements on the ancillas, can be exploited to perform the desired POVM.

\begin{figure*}[ht!]
    \includegraphics[width=1\textwidth]{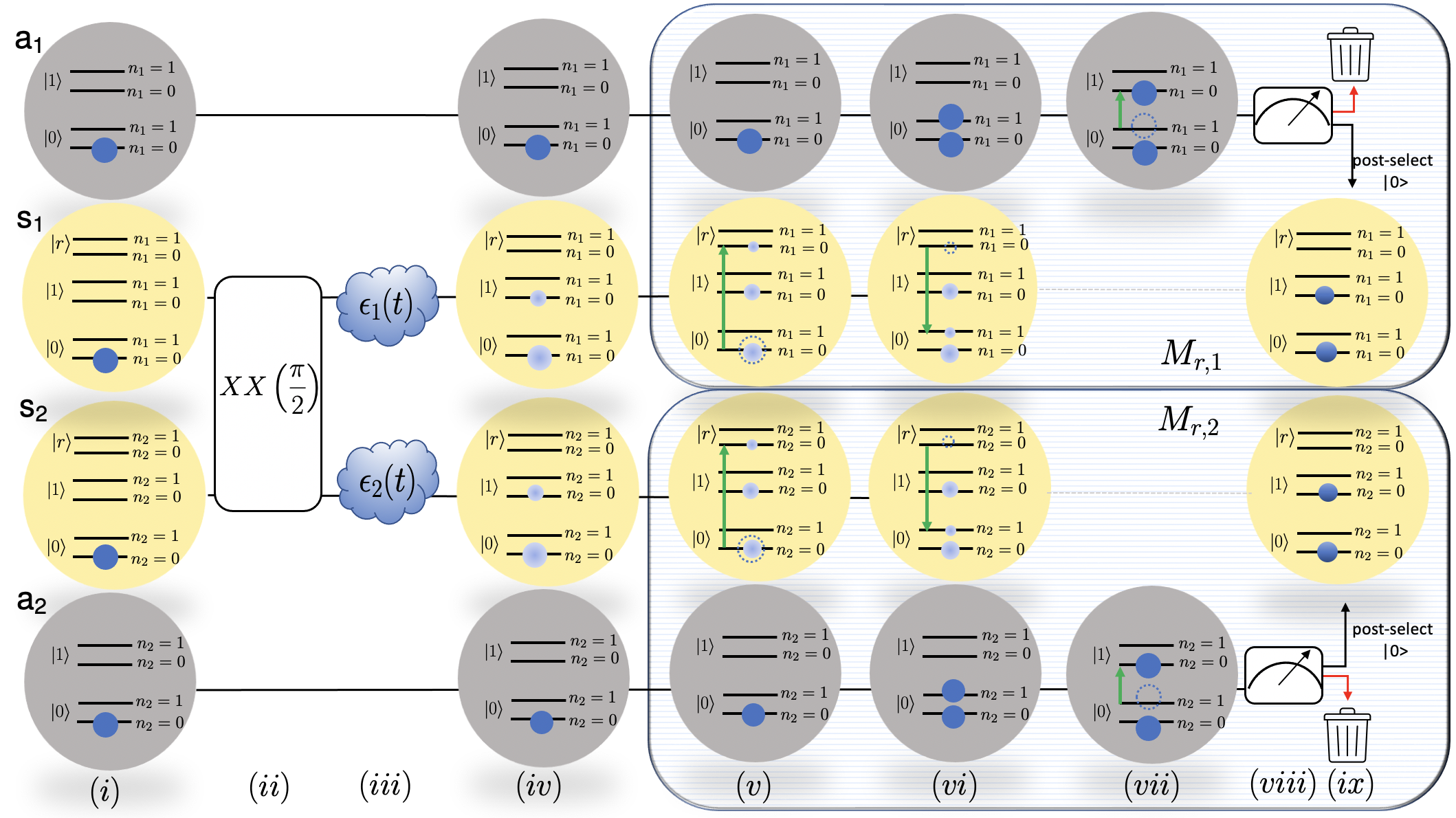}\\
    \caption{{\bf Amplitude-damping reversal by quantum logic spectroscopy:} The system ${\rm s_1,s_2}$ and ancillary ${\rm a_1,a_2}$ subsystems of Fig.~\ref{fig:general_filtering_scheme} are composed of both internal and motional levels. In the left-hand side, we include the two lowest Fock states for the vibrational ladder of states associated to each of the system and ancilla qubits. Note that the vibrational Fock levels of ${\rm s_1,a_1}$ (and of ${\rm s_2,a_2}$) correspond to the same common mode.
    $(i)$ The initial state 
    has no internal/motional excitation. Using the trapped-ion native gates, $(ii)$ an entangled pair is generated by the entangling gate $XX(\pi/2)$~\eqref{eq:MS_x}, which is followed by $(iii)$ the amplitude-damping channels $\epsilon_{1}(t),\epsilon_{2}(t)$~\eqref{eq:AD_channel}, depicted by two clouds, that act during a time  $t$ that sets the error rate~\eqref{eq:p_T1}. The resulting effect, summarized in $(iv)$, is to shuffle the amplitudes, as depicted by the dashed blue circles  of the subsequent column. The reversal/filtering operations that are then carried out to distill a better entangled state and reverse the environmental noise are depicted inside the following two boxes, which correspond to the sequence of carrier $(v)$ and sideband pulses $[vi-vii]$, followed by $(viii)$ a projective measurement, and $(ix)$ post-selection on the ancillas. As discussed in the text, by adjusting the duration of the pulses~\eqref{eq:timing_carrier}, and post-selecting on the measurement outcomes~\eqref{eq:measurement}, one obtains the desired non-unitary filter~\eqref{eq:reversal_operations}. }
    \label{fig:general_filtering_first_scheme} 
\end{figure*}

\subsection{Scheme A: Asymmetric POVMs via quantum logic spectroscopy}\label{sec:schemeA}

The scheme to realize an asymmetric POVM consists of a series of  unitary operations applied to the  system and ancilla ions, followed by a final projective measurement   on the ancilla qubits that post-selects certain outcomes.
The underlying idea is  similar to that of  quantum logic spectroscopy (QLS)~\cite{Schmidt749},  
in which information stored in  the system qubits can be coherently transferred onto the ancilla ions through  
the common vibrational modes provided by the Coulomb interaction between the ions.
The procedure ends with a projective measurement of the ancilla qubits that induces a POVM on the system qubits. 
The optimal implementation would use ancilla ions from a different atomic species or isotope, which  reduces  the effect  
that light scattering during a fluorescence-enabled measurement of the ancillas can have on the system qubits. Alternatively, 
ion shuttling could be used to transport the ancilla ions to a measurement trapping region located far away from the system qubits~\cite{Kielpinski2002,shuttling}. 
We note that another possibility to implement an asymmetric POVM 
without ancillas is to use more internal states of the ions, as discussed in \cite{PhysRevLett.111.180501, PhysRevA.102.022426}.

The initial stage of construction of our asymmetric POVMs proceeds by analogy to the symmetric case~\cite{PhysRevA.87.012131} and thus exploits three different levels from the atomic level structure of the ions. We denote these by $\{\ket{0}\s{s_{q}}, \ket{1}\s{s_{q}}, \ket{r}\s{s_{q}}\}$ for the  ions $\rm{q}=\{1,2\}$ that store the two system/data qubits (see Fig.~\ref{fig:general_filtering_scheme}).
The state $\ket{r}\s{s_{q}}$ is an auxiliary metastable excited state that connects the qubits states in a lambda configuration.
The two ancillary subsystems ${\rm a}_1,{\rm a_2}$ also shown in Fig.~\ref{fig:general_filtering_scheme} are provided by additional ions, each of which  contributes with a pair of internal levels, which are  initially prepared in  the state $\rho\s{a}(t_{0})=\ketbra{0}\s{a_{1}}\!\!\otimes\ketbra{0}\s{a_{2}}$.
The system  and  ancilla ions form a linear chain aligned along the null of the rf-field of a Paul trap that confines the ion register. Accordingly, their Coulomb interaction couples the small vibrations around the ion crystal equilibrium position, giving rise to the common vibrational modes. We  select two specific vibrational modes, $\rm{m}_{1}$ and $\rm{m}_{2}$, each of which describes specific collective vibrations along a particular axis direction $k_{1}$ and $k_{2}$. Assuming initial laser-cooling conditions to the ground state of both modes, the motional state can be described by the tensor product of two Fock states with no vibrational quanta, $\rho\s{m}(t_{0})=\ketbra{0}\s{m_{1}} \otimes\ketbra{0}\s{m_{2}}$.  The initial state of the scheme is then written as the tensor product state
$\rho(t_{0})=\rho\s{s}(t_{0})\otimes\rho\s{a}(t_{0})\otimes\rho\s{m}(t_{0})$, where $\rho\s{s}(t_{0})$ represents the  internal state of the system qubits that, after being subjected to the entangling gate  depicted in Fig.~\ref{fig:general_filtering_first_scheme}, is close to one of the four maximally-entangled pairs, i.e., the four Bell states.
 
We assume that the remaining vibrational modes do not intervene in the protocol and thus act as mere spectators. We also assume that the heating on each of the vibrational modes is  vanishingly small.
Using a trapped-ion native gate set, we can remove the Hadamard gate in Fig.~\ref{fig:general_filtering_scheme} by using a M\o lmer-S\o rensen (MS) gate~\cite{PhysRevLett.82.1835,PhysRevLett.82.1971} instead of the CNOT gate, specified by
\beq
\label{eq:MS_x}
U_{\rm id}=XX\left(\frac{\pi}{2}\right)=\frac{1}{\sqrt{2}}(\mathbb{I}-\ii\sigma^{x}\s{s_{1}}\sigma^{x}\s{s_{2}}).
\eeq
This entangling gate, represented by $XX(\pi/2)$  in Fig.~\ref{fig:general_filtering_first_scheme}, readily generates four maximally-entangled pairs on the system qubits that are locally equivalent to the Bell states of Eq.~\eqref{eq:bell_pairs}. In the left-hand side of this figure, step $(i)$ represents an initial state with the system and ancilla qubits initialised in  $\ket{0}$. Note that we also draw the two lowest vibrational Fock levels for the corresponding common modes, which will be used in subsequent steps. In step $(ii)$, the MS gate produces the entangled pair $(\ket{0}_{{\rm s}_1}\!\!\otimes\ket{0}_{{\rm s}_2}-\ii\ket{1}_{{\rm s}_1}\!\!\otimes\ket{1}_{{\rm s}_2})/\sqrt{2}$ for the system qubits.  In step $(iii)$, this unitary is followed by the amplitude decay channel described by Eq.~\eqref{eq:damping_composition}, leading to a partially-entangled mixed state $\rho_{\rm id}=U_{\rm id}\rho(t_0)U_{\rm id}^\dagger\mapsto\tilde{\rho}_{\rm id}$.  As a consequence of spontaneous emission, the probabilities to find the system qubits in either of the two possible states will now no longer be equal, which is depicted in step $(iv)$ by the different size of the shaded blue balls of the corresponding levels.

We now show how to construct an  asymmetric POVM that can reverse the amplitude damping. This requires  a sequence of gates  to implement $M_{r,q}$ in Eq.~\eqref{eq:reversal_operations}, which is depicted inside the two rectangular boxes on the right of Fig.~\ref{fig:general_filtering_first_scheme}. The scheme starts with three consecutive unitary operations, steps $(v)$ - $(vii)$. The first one, step $(v)$, is $O_{1}=R^{c}\s{s_{1}}(\theta_{1}, \phi_{1})\otimes R^{c}\s{s_{2}}(\theta_{1}, \phi_{1})$, and  consists of a carrier pulse on each system ion in resonance with the transition $\ket{0}\s{s_{q}} \rightarrow \ket{r}\s{s_{q}}$~\cite{doi:10.1080/00107514.2011.603578}. This reads
\begin{equation}
\label{eq:carrier_unitary}
    R\s{s_{q}}^{\rm c}(\theta_{1}, \phi_{1})=\exp\left\{\ii\frac{\theta_{1}}{2}(e^{\ii\phi_{1}}\sigma\s{s_{q}}^{+}+e^{-\ii\phi_{1}}\sigma\s{s_{q}}^{-})\right\},
\end{equation}
 where $\sigma\s{s_{q}}^{+}=\ket{r}\bra{1}\s{s_{q}}(\sigma^{-}\s{s_{q}}=\ket{1}\bra{r}\s{s_{q}})$ are the spin raising (lowering) operators. This carrier pulse, represented by the green arrows of the first carrier pulse in the boxes of Fig.~\ref{fig:general_filtering_first_scheme}, must  act for a specific  duration
\beq
\label{eq:timing_carrier}
t_{1}=\frac{2}{\Omega^{c}_{1}}\cos^{-1}(\sqrt{1-p_{r}}).
\eeq
We set  $\phi_{1}=0$ in Eq.~\eqref{eq:carrier_unitary}, and define the pulse area $\theta_{1}=\Omega_{1}^{c}t_{1}$ in terms of the carrier Rabi frequency $\Omega_{1}^{c}$, which  depends on a parameter  $p_{r}$ that controls the strength of the  reversal operation. In Fig.~\ref{fig:general_filtering_first_scheme}, the effect of this carrier is depicted by a partial transfer of the shaded blue amplitude to the auxiliary $r$ level.

In step $(vi)$  we then apply a red-sideband pulse  $O_{2}=R^{\rm rsb}\s{s_{1}, m_{1}}(\theta_{2}, \phi_{2})\otimes R^{\rm rsb}\s{s_{2}, m_{2}}(\theta_{2}, \phi_{2})$ to each of the system ions \cite{doi:10.1080/00107514.2011.603578}. This corresponds to the unitary 
 \begin{equation}
  \label{eq:red_sideband}
   R\s{s_{q}, m_{q}}^{\rm rsb}(\theta_{2}, \phi_{2})=\exp\left\{\ii\frac{\theta_{2}}{2}(e^{\ii\phi_{2}}\sigma\s{s_{q}}^{+}a\s{m_{q}}+e^{-\ii\phi_{2}}\sigma\s{s_{q}}^{-}a^{\dagger}\s{m_{q}})\right\},  
 \end{equation}
 where $a\s{m_{q}}(a^{\dagger}\s{m_{q}})$ are the annihilation (creation) operators of  phonons in the common vibrational mode $\rm{m}_{q}$. This sideband is resonant with the transition $\ket{r}\s{s_{q}}\otimes\ket{n}\s{m_{q}}\rightarrow \ket{0}\otimes\s{s_{q}}\ket{n+1}\s{m_{q}}$, which increases the phonon number of the corresponding vibrational Fock state. To map the relevant information into the common mode,  we set  the phase to  $\phi_{2}=0$, and the pulse area to $\theta_{2}=\pi$, such that the pulse duration is
 \beq
t_{2}=\frac{\pi}{\Omega_{2}^{\rm rsb}}.
 \eeq 
 Here, we use the  definition of the red-sideband Rabi frequency $\Omega_{2}^{\rm rsb}=\eta\Omega_{2}^{c}$,

where $\Omega_{2}^{c}$ is the carrier Rabi frequency of this second pulse, 
$\eta=(\hbar {k}_L/2M\s{m}\omega\s{m_{q}})^{1/2}$ 
is the Lamb-Dicke parameter defined in terms of the frequency of the $\rm{m}$-th vibrational mode $\omega\s{m_{q}}$, and the mass of the  ions $M\s{m}$ determines the contribution of the ion $q$ to the specific normal mode. 

 In Fig.~\ref{fig:general_filtering_first_scheme}, the effect of this sideband pulse is depicted by green lines pointing down, which  partially transfer  the shaded blue amplitude of the  $r$ level onto the vibrational Fock state.

At this stage, we  make use of the ancillary qubits by mapping the amplitude of the common  vibrational Fock state onto the internal states of the ancilla qubits.
For the third operation in step $(vii)$, $O_{3}=R\s{a_1,m_1}^{\rm rsb}(\theta_{3}, \phi_{3})\otimes R\s{a_2,m_2}^{\rm rsb}(\theta_{3}, \phi_{3})$, we  thus apply another red-sideband pulse to each ancilla qubit
 \begin{equation}
 \label{eq:red_sideband_anc}
   R\s{a_{q},m_{q}}^{\rm rsb}(\theta_{3}, \phi_{3})=\exp\left\{\ii\frac{\theta_{3}}{2}(e^{\ii\phi_{3}}\sigma\s{a_{q}}^{+}a\s{m_{q}}+e^{-\ii\phi_{3}}\sigma\s{a_{q}}^{-}a\s{m_{q}}^{\dagger})\right\}.
 \end{equation}

In this case, the sideband is  in resonance with the transition $\ket{0}\s{a_{q}}\ket{n+1}\s{m_{q}}\rightarrow \ket{1}\s{a_{q}}\ket{n}\s{m_{q}}$, and we set the phase and pulse area to $\phi_{3}=0$ and $\theta_{3}=\pi$, respectively. Accordingly, the time duration is again
 \beq
t_{3}=\frac{\pi}{\Omega_{3}^{\rm rsb}},
 \eeq 
 where all parameters are defined by analogy with the previous red-sideband pulse.

Following the scheme of Fig.~\ref{fig:general_filtering_first_scheme}, this operation acts on the ancillary subspaces depicted by grey circles inside the two boxes, where the green lines point upwards and denote how the vibrational Fock excitation is converted into an excitation of the ancilla qubits.

The final step  $(viii)$, which is crucial to engineering the POVM and the non-unitary filter, consists of a projective measurement on each of the ancilla qubits. This is followed by post-selecting in step $(ix)$ on the outcomes that are consistent with the state $\ket{0}\s{a_{q}}$. Using the  projectors
\beq
E^{\sigma\s{a_{q}}}_\pm=(\mathbb{I} \pm \sigma\s{a_{q}}^{z}  )/2,
\eeq

where $\sigma\s{a_{q}}^{z}=\ketbra{0}\s{a_{q}}-\ketbra{1}\s{a_{q}}$ for $\rm{q}=\{1,2\}$,
the reduced density matrix for the system  is equivalent to a POVM measurement with the four possible outcomes 
\beq
\label{eq:measurement}
\tilde{\rho}\s{id}\to \left\{\begin{matrix}
M_{r,2}^{\phantom{\dagger}}M_{r,1}^{\phantom{\dagger}}\tilde{\rho}\s{id} M_{r,1}^{\phantom{\dagger}}M_{r,2}^{\phantom{\dagger}} ,\hspace{1ex} {\rm if\,\,ancillas \,\,in\,\, \ket{0}\s{a_1}\otimes\ket{0}\s{a_2},}\\ 
\overline{M}_{r,2}^{\phantom{\dagger}}{M}_{r,1}^{\phantom{\dagger}}\tilde{\rho}\s{id} {M}_{r,1}^{\phantom{\dagger}}\overline{M}_{r,2}^{\phantom{\dagger}},\hspace{1ex} {\rm if\,\, ancillas \,\, in\,\,\ket{0}\s{a_1}\otimes\ket{1}\s{a_2},}\\
{M}_{r,2}^{\phantom{\dagger}}\overline{M}_{r,1}^{\phantom{\dagger}}\tilde{\rho}\s{id} \overline{M}_{r,1}^{\phantom{\dagger}}{M}_{r,2}^{\phantom{\dagger}},\hspace{1ex} {\rm if\,\, ancillas \,\, in\,\,\ket{1}\s{a_1}\otimes\ket{0}\s{a_2},}\\
\overline{M}_{r,2}^{\phantom{\dagger}}\overline{M}_{r,1}^{\phantom{\dagger}}\tilde{\rho}\s{id} \overline{M}_{r,1}^{\phantom{\dagger}}\overline{M}_{r,2}^{\phantom{\dagger}},\hspace{1ex} {\rm if\,\, ancillas \,\, in\,\,\ket{1}\s{a_1}\otimes\ket{1}\s{a_2}.}
\end{matrix}
\right.
\eeq
Here,  we have introduced the  following operators $M_{r,1}=M_r\otimes\mathbb{I},M_{r,2}=\mathbb{I}\otimes M_r$, and similarly for $\overline{M}_{r,1},\overline{M}_{r,2}$, where 
\beq
\label{eq:POVM_WM}
M_{r}=\frac{1}{\sqrt{P_r}}\begin{pmatrix}
\sqrt{\bar{p}_{r}} & 0 \\
0 & 1 \\
\end{pmatrix},\hspace{1ex} \overline{M}_{r}=\frac{1}{\sqrt{\overline{P}_r}}\begin{pmatrix}
\sqrt{{p}_{r}} & 0 \\
0 & 1 \\
\end{pmatrix},
\eeq
and the probability for the  outcomes is determined by  $P_{r}={\rm Tr}\{M_{r,i}^{\dagger}M_{r,i}\rho_{\rm s}\}$, with $\overline{P}_{r}=1-{P}_{r}$. These POVMs resemble the Kraus operators for the asymmetric amplitude damping channel of Eq. (\ref{eq:Krauss}) \and are also manifestly asymmetric, since 
in the limit $p_r\to 1$, the POVM operator $M_{r}$ maps onto a von Neumann projection, $M_{r}\to\ket{1}\bra{1}$, which connects the reversal operations with an infinitely sharp, i.e., a strong projective measurement onto a single qubit state, whilst for $p_r\to 0$ the POVM operator $M_{r}$ maps to the identity matrix, which has no effect on the qubit states.
These POVMs thus represent an asymmetric version of a weak measurement, in contrast to the symmmetric POVMs proposed for QLS in~\cite{PhysRevA.87.012131}.

It is now a straightforward to see how this asymmetric POVM can be used to reverse the asymmetric amplitude-damping channel. By post-selecting on those measurement outcomes consistent with the $\ket{0}\s{a_q}$ state, we are effectively applying the projection $O_{4}=E^{\sigma_{{\rm a}_1}}_+E^{\sigma_{{\rm a}_2}}_+$, and thereby introducing the non-unitary character of the filtering process. The output density matrix after all these  consecutive steps reads
\begin{equation}
\begin{split}
     \rho_{\rm f}&=M_{r,2}^{\phantom{\dagger}}M_{r,1}^{\phantom{\dagger}}\tilde{{\rho}}_{\rm id} M_{r,1}^{\phantom{\dagger}}M_{r,2}^{\phantom{\dagger}},
\end{split}
\end{equation}
in accordance with Eq.~\eqref{eq:reversal_operations}. Note that the reversal parameter $p_r$ can be fully controlled by modifying the timing of the first carrier pulse~\eqref{eq:timing_carrier}. If we look at the effect of the original amplitude-damping channel with Kraus operators~(\ref{eq:Krauss}), it is clear that 
the reversal operator $M_{r}$ in Eq.~\eqref{eq:POVM_WM} can invert most of the effect of the amplitude damping if one controls the time duration of the carrier pulse such that 
\beq
\label{eq:reversal_T1}
p_{r}=p=1-\ee^{-t/T_1}.
\eeq

Essentially, the non-unitary filter 
reduces the amplitude of the $\ket{0}\s{s_{q}}$ system qubit states  to increase the fidelity with the target entangled state, leading to a single-copy probabilistic quasi-distillation. The practical observation is that the prior information required for this quasi-distillation is no longer related to the amplitudes of the initial entangled state, as in the Procrustean method~\cite{PhysRevA.53.2046,GISIN1996151}, but  instead depends on the $T_1$ time of the noise channel~\eqref{eq:reversal_T1} and can serve to protect any maximmaly-entangled pair. Let us also emphasize that each of the blocks of Fig.~\ref{fig:general_filtering_first_scheme}, which lead to the corresponding filters $M_{r,1}$ and $M_{r,2}$ are only composed of LOCC operations. The two partially-entangled qubits ${\rm s_1,s_2}$ can  thus be  spatially separated, and these LOCC operators serve to distill a state with a larger  fidelity with the target maximally-entangled state, as will be discussed with specific numerical simulations in the following section.

\begin{figure*}[ht!]
    \includegraphics[width=1.6\columnwidth]{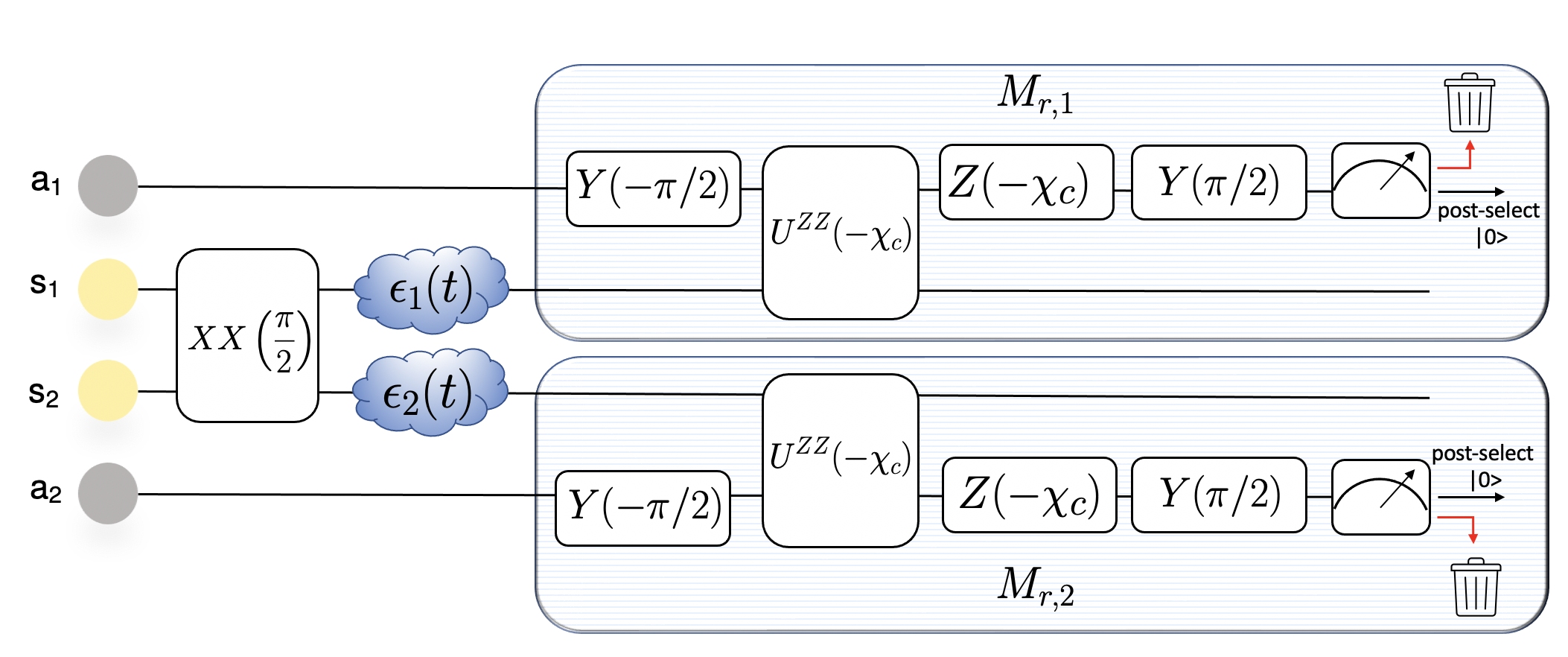}\\
    \caption{{\bf Amplitude-damping reversal by system-ancilla entangling gates:} In this scheme, the two blocks that realize the reversal operations $M_{r,1},M_{r,2}$ are modified from those in Fig.~\ref{fig:general_filtering_first_scheme}. Here, we perform a sequence of rotations on the ancilla qubits $Y(\pm\pi/2)=\ee^{\pm\ii\pi\sigma^y\s{a_q}/4}$, and $Z(\pm\pi/2)=\ee^{\ii\chi_{\rm c}\sigma^z\s{a_q}/2}$, together with an entangling  geometric phase gate $U^{ZZ}(-\chi_{\rm c})=\ee^{\ii\chi_{\rm c}\sigma^z\s{a_q}\sigma^z\s{s_q}}$. The specific ordering is depicted in the two shaded boxes, which also include the ancilla projective  measurement and post-selection. }
    \label{fig:general_filtering_second_scheme} 
\end{figure*}

\subsection{Scheme B: Asymmetric POVMs via entangling gates}\label{sec:schemeB}

In this subsection, we introduce a different scheme that still requires using two ancilla qubits, but  no longer relies on  motional Fock states with a single motional excitation.  Instead, it exploits  two-qubit gates between system and ancilla qubits, which are mediated by off-resonant excitations of the common motional modes in a way that is considerably insensitive to the thermal populations~\cite{doi:10.1080/00107514.2011.603578}. 
The scheme is summarized in Fig.~\ref{fig:general_filtering_second_scheme}.
The full unitary part of the scheme  can be described by the following operator
\begin{equation}\label{eq:unitary_schemeB}
    U\s{q}=e^{\ii\frac{\pi}{4}\sigma^{y}\s{a_{q}}}e^{\ii \frac{\chi_{c}}{2}\sigma^{z}\s{a_{q}}}e^{\ii \frac{\chi_{c}}{2}\sigma^{z}\s{a_{q}}\sigma^{z}\s{s_{q}}}e^{-\ii\frac{\pi}{4}\sigma^{y}\s{a_{q}}},
\end{equation}
which must be applied to each separate pair ${\rm s_q,a_q}$ for ${\rm q=\{1,2\}}$. This unitary
 combines single-qubit rotations on the ancilla  qubits with an entangling geometric phase gate~\cite{Leibfried2003} between the ancilla and the corresponding system qubit. We have found that in order to implement the asymmetric POVMs~\eqref{eq:reversal_operations},  the geometric phase of the later must be   
\beq 
\chi_{c}=\cos^{-1}(\sqrt{1-p_{r}}),
\eeq
which is the analogue of Eq.~\eqref{eq:timing_carrier} in the previous QLS-based scheme.

Following these operations, as depicted in Fig.~\ref{fig:general_filtering_second_scheme}, the final step consists of a projective measurement and post-selection on the ancilla qubit in state $\ket{0}\s{a_q}$ as in the previous scheme. Finally, the output density matrix after these four consecutive steps can be written as 
\begin{equation}
\begin{split}
     \rho_{\rm f}&=M_{r,2}^{\phantom{\dagger}}M_{r,1}^{\phantom{\dagger}}\tilde{\rho}_{\id} M_{r,1}^{\phantom{\dagger}}M_{r,2}^{\phantom{\dagger}}
\end{split}
\end{equation}
where we have introduced 
$M_{r,{\rm q}}=E^{\sigma\s{a_q}}_+U\s{q}/\sqrt{P_r}$ in terms of the previous ancilla projectors $E^{\sigma\s{a_q}}_+$,
with the corresponding post-selection  probabilities $P_{r}={\rm Tr}\{M_{r,{\rm q}}^{\dagger}M_{r,{\rm q}}\tilde{\rho}_{\rm id}\}$. 

In the following section, we will show numerically that this scheme also implements the desired non-unitary filter, and that it is more robust than the scheme of Sec.~\ref{sec:schemeA} to thermal fluctuations of the vibrational mode that is used to mediate the geometric phase gate. \\

\section{results}\label{sec:results}

So far, we have presented the single-copy quasi-distillation protocol, and discussed two possible schemes to realise the protocol in trapped-ion platforms. We  have also illustrated how it can improve the fidelity of a specific mixed state resulting from amplitude-damping noise~\eqref{eq:infidelity_filtered}. As remarked already, an interesting property of the filtering scheme is that it can work for any target entangled state. In order to show that this is indeed the case, in this section we
 present analytical and numerical results showing that  both schemes lead to the desired single-copy quasi-distillation,  and also address some possible practical limitations. The criterion to determine the success of the non-unitary probabilistic filtering to protect any entangled state against amplitude damping is  dictated by the increase of the average gate fidelity of the noisy implementation of the unitary $U_{\rm id}$. 
 
 In Sec.~\ref{sec:ideal} we now present analytical formulas when the whole filtering protocol is executed under ideal conditions and numerical results validating the performance of the protocol as a tool for suppression of errors due to amplitude damping. In Secs.~\ref{sec:limits_A} and \ref{sec:limits_B} we then show  numerical results for more realistic situations that could limit the performance of the protocol for each scheme.

\subsection{Ideal implementation of the filtering schemes}\label{sec:ideal}
To compare an ideal unitary operation with its actual implementation due to  noise and  experimental imperfections, we use the measure provided by the average gate fidelity \cite{nielsen00}, namely
\beq\label{eq:avg_Fg}
\bar{\mathcal{F}}_{g}(U_{\rm id}, \epsilon)=\int d \Psi_{\rm 0}\bra{\Psi_{\rm 0}} U_{\rm id}^{\dagger} \epsilon(\tilde{\rho}_{\rm id})U_{\rm id}\ket{\Psi_{\rm 0}}.
\eeq
Here $U_{\rm id}$ is the target unitary  and $\epsilon(\tilde{\rho}_{\rm id})$ represents the evolution of the system under imperfect implementations of the unitary. Complete preservation of the quantum information implies $\bar{\mathcal{F}}_{g}(U_{\rm id}, \epsilon)=1$, which corresponds  to a perfect implementation of the unitary with $\epsilon=\mathbb{I}$, i.e., zero noise.

The integral in Eq.~\eqref{eq:avg_Fg} must be performed over the Hilbert space of all possible initial  two-qubit states $\Psi_{\rm 0}$.
Alternatively, one can estimate $\bar{\mathcal{F}}_{g}(U_{\rm id}, \epsilon)$ by the entanglement fidelity~\cite{PhysRevA.54.2614,PhysRevA.60.1888,NIELSEN2002249}, which is defined for a single initial state $\ket{\phi_{m}}$ that is a maximally entangled state of the system with an auxiliary quantum system, according to

\beq
\label{eq:F_e_definition}
 \bar{\mathcal{F}}_{e}(U_{\rm id}, \epsilon)=\bra{\phi_{m}}\mathbb{I}_{d}\otimes U_{\rm id}^{\dagger}\epsilon(\ket{\phi_{m}}\bra{\phi_{m}})\mathbb{I}_{d}\otimes U_{\rm id}\ket{\phi_{m}}.
\eeq
Here $U_{\rm id}$ is an $N$ qubit (or qudit) unitary acting on the system alone. 
This measure makes use of an initial state $\ket{\phi_{m}}=\sum_{\alpha=1}^{d}\ket{\alpha}\otimes\ket{\alpha}/\sqrt{d}$ 
that is  maximally entangled  between two subsystems. In our case these are firstly the $N=2$ data qubits where the information is encoded, and secondly an auxiliary system with two spectator qubits.  The dimension of the $N$-qubit unitary is then $d=2^{N}=4$.
Both the unitary $\mathbb{I}_{d}\otimes U_{\rm id}$, and the noise channel $\epsilon(\ket{\phi_{m}}\bra{\phi_{m}})=\mathbb{I}_{d}\otimes\epsilon(\tilde{\rho}_{\rm id})$, only act on the data qubits  whilst the spectator qubits  remain unaffected.

Remarkably, one can derive a simple  formula  connecting the entanglement fidelity with the gate fidelity~\cite{PhysRevA.60.1888,NIELSEN2002249}, namely
\begin{equation}\label{eq:avg_Fg_and_Fe}
    \begin{split}
       \bar{\mathcal{F}}_{\rm g}(U_{\rm id}, \epsilon)=\frac{d\bar{\mathcal{F}}_{e}(U_{\rm id}, \epsilon)+1}{d+1}.
    \end{split}
\end{equation}
Since, $\bar{\mathcal{F}}_{g}\geq \bar{\mathcal{F}}_{e}$ \cite{PhysRevA.54.2614}, the entanglement fidelity represents a lower bound for the average gate fidelity.\\

We now derive analytical expressions for the entanglement fidelity of our protocol in both the  absence and presence of the noise-filtering operations. 
The unitary gate we consider is the two-qubit entangling gate $U_{\rm id}=XX\left(\frac{\pi}{2}\right)$, Eq. (\ref{eq:MS_x}), and the noise evolution $\epsilon(\tilde{\rho}_{\rm id})$ is given in terms of the Kraus operators for the uncorrelated noise channels of Eqs.~(\ref{eq:damping_composition}) and (\ref{eq:AD_channel}).
We choose $\ket{\alpha} \in \{ \ket{00}, \ket{01}, \ket{10}, \ket{11}\}$ for the set of operator basis states. Starting from Eq.~(\ref{eq:F_e_definition}), the entanglement fidelity for the unfiltered evolution reduces to
\begin{equation}\label{eq:F_entanglement_unf}
    \begin{split}
        \bar{\mathcal{F}}_{e}^{\rm unf}(U_{\rm id}, \epsilon)=\frac{1}{4d^{2}}\left|\sum_{i,j=0}^{1}{\rm{Tr}}\big\{ K_{i,{\rm s}_1}\otimes K_{j,{\rm s}_2}\big\}\right|^{2},
    \end{split}
\end{equation}
with $d=4$ for $N=2$  system qubits, where we have made use of the analysis in~\cite{Johnston_2011} to write the result in terms of the Kraus operators.

Including the reversal/filtering operations to distill a single-copy entangled state, and applying $M_{r}=M_{r,1}\otimes M_{r,2}$ in Eq.~(\ref{eq:reversal_operations}) via either scheme A (Sec. \ref{sec:schemeA}) or scheme B (Sec. \ref{sec:schemeB}), modifies the entanglement fidelity to

\begin{equation}\label{eq:F_entanglement_f}
    \begin{split}
        \bar{\mathcal{F}}_{e}^{\rm f}(U_{\rm id}, \epsilon)=\frac{1}{4d^{2}P_{r}}\left|\sum_{i,j=0}^{1}{\rm{Tr}}\big\{ M_{r}(K_{i,{\rm s}_1}\otimes K_{j,{\rm s}_2})\big\}\right|^{2}.
    \end{split}
\end{equation}
Here  $P_{r}$ is  the probability of success of the reversing operation, 
given by $P_{r}={\rm{Tr}}\{M_{r}^{\dagger}M_{r}\epsilon(\ket{\phi_{m}}\bra{\phi_{m}})\}$.

The implementation of $M_{r}$ and hence the sequence of gates employed for realization of the protocol differs between the QLS-based and entangling based schemes presented in Section~\ref{sec:trapped_ions}. In scheme A (Fig.~\ref{fig:general_filtering_first_scheme}) $M_{r}$ takes the form 
\begin{equation}\label{eq:F_entanglement_f_A}
    \begin{split}
         M_{r}^{A}= M_{r,1}^{A}\otimes M_{r,2}^{A}=O_{4}O_{3}O_{2}O_{1}
    \end{split}
\end{equation}
where the various operations $O_{i}$ correspond to the set of unitaries depicted inside the box of Fig.~\ref{fig:general_filtering_first_scheme}, applied to each of the two parties. In particular,  we use the carrier pulses $O_{1}=R^{c}\s{s_{1}}(\theta_{1}, \phi_{1})\otimes R^{c}\s{s_{2}}(\theta_{1}, \phi_{1})$ in Eq.~(\ref{eq:carrier_unitary}), the sidebands  $O_{2}=R^{\rm rsb}\s{s_{1}, m_{1}}(\theta_{2}, \phi_{2})\otimes R^{\rm rsb}\s{s_{2}, m_{2}}(\theta_{2}, \phi_{2})$
in Eq.~(\ref{eq:red_sideband}), and $O_{3}=R\s{a_1,m_1}^{\rm rsb}(\theta_{3}, \phi_{3})\otimes R\s{a_2,m_2}^{\rm rsb}(\theta_{3} , \phi_{3})$ 
in Eq.~(\ref{eq:red_sideband_anc}). Finally,  $O_{4}=E^{\sigma_{{\rm a}_1}}_+E^{\sigma_{{\rm a}_2}}_+$
are the projectors  $E^{\sigma\s{a_{q}}}_\pm=(\mathbb{I} \pm \sigma\s{a_{q}}^{z}  )/2$ onto the $\ket{0}\s{a_{q}}$ states for both ancilla qubits $\rm{a_{1}}$ and $\rm{a_{2}}$, that are used for post-selection.\\
For scheme B (Fig.~\ref{fig:general_filtering_second_scheme}), the  reversal operation is instead implemented by the sequence of unitaries
\begin{equation}\label{eq:F_entanglement_f_B}
    \begin{split}
         M_{r}^{B}= M_{r,1}^{B}\otimes M_{r,2}^{B}=U_{1}\otimes U_{2},
    \end{split}
\end{equation}
where $U_{1}$ and $U_{2}$ are described by the sequence of single and two-qubit gates  given in Eq.~(\ref{eq:unitary_schemeB}), which involve the ancilla-system qubit pairs, $\{\rm{a_1},\rm{s_1}\}$ and $\{\rm{a_2},\rm{s_2}\}$, respectively. 

\begin{figure}[ht!]
    \includegraphics[width=1\columnwidth]{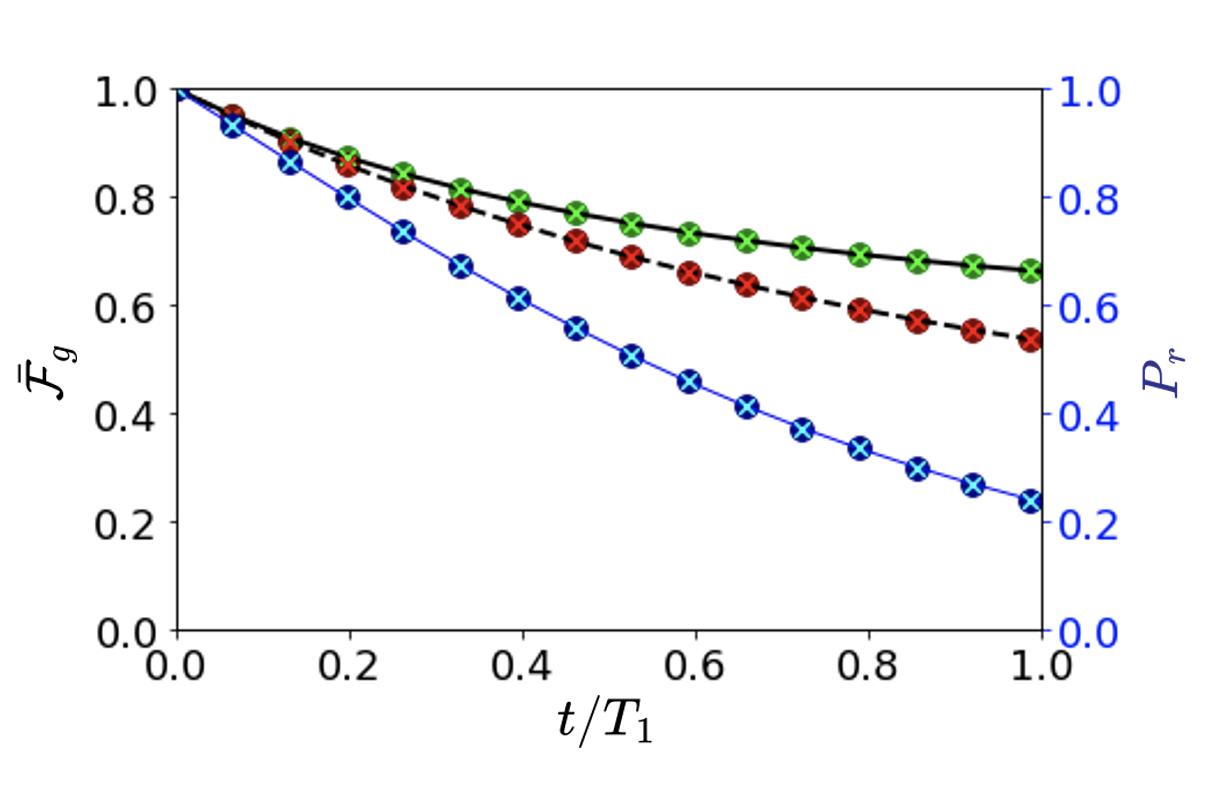}\\
    \caption{{\bf Average gate fidelities and success probability for ideal case implementation of error reversal schemes A and B:} Plotted lines correspond to the analytical formulas from Sec.\ref{sec:ideal} as specified below. Solid circles and cross symbols represent the results of the numerical simulations of the process following scheme A and scheme B, respectively.  The black font y-axis on the left represents the filtered average gate fidelity and the blue y-axis on the right represents the success probability $P_r$. The x-axis represents time in units of the amplitude damping decay time $T_{1}$. $T\s{1}=0.8$ s for the results shown here.
    The black solid line is the expression for the analytical average gate fidelity $\bar{\mathcal{F}}_{g}^{\rm f}$ (\ref{eq:F_avg_f}) when the filtering protocol is implemented after the unitary. Scattered green dots (lime crosses) represent the numerical simulations of the average gate fidelities for scheme A(B), $\bar{\mathcal{F}}_{g}^{\rm f, A}$ ($\bar{\mathcal{F}}_{g}^{\rm f, B}$).
    The dashed black line is the bare average gate fidelity $\bar{\mathcal{F}}_{g}^{\rm unf}$ (\ref{eq:F_avg_unf}). Scattered maroon dots (red crosses) are the simulations of the unfiltered average gate fidelity for scheme A(B), $\bar{\mathcal{F}}_{g}^{\rm unf, A}$ ($\bar{\mathcal{F}}_{g}^{\rm unf, B}$).
    The blue solid line is the success probability of the process given by $P_{r}={\rm{Tr}}\{M_{r}^{\dagger}M_{r}\epsilon(\ket{\phi_{m}}\bra{\phi_{m}})\}$. Scattered blue dots (cyan crosses) are the simulations of the unfiltered average gate fidelity for scheme A(B), $P_{r,A}$ ($P_{r,B}$). The measurement reversal operation , $M_{r}$, takes the form of Eq.(\ref{eq:F_entanglement_f_A}) and Eq.(\ref{eq:F_entanglement_f_B}), for scheme A and B, respectively.}
    \label{fig:ideal_case_F_g} 
\end{figure}

For the ideal case scenario where the gates have no errors and  one assumes an initial motional ground state, numerical simulation of the performance of both non-unitary filters against amplitude damping errors leads to the same result, namely  $M_{r,j}^{A}=M_{r,j}^{B}=M_{r,j}$ in Eq.~\eqref{eq:reversal_operations}.  This is confirmed by numerical simulations of the protocol of the scheme A and scheme B using a full-density matrix formalism.
Therefore, in the ideal case both schemes yield the same entanglement fidelity, $\bar{\mathcal{F}}_{e}^{\rm f}=\bar{\mathcal{F}}_{e}^{\rm f, A}=\bar{\mathcal{F}}_{e}^{\rm f, B}$, as is also evident from Eq.(\ref{eq:F_entanglement_unf}). 
Using Eq.(\ref{eq:avg_Fg_and_Fe}) then implies identical average gate fidelities, which is verified in the numerical results plotted in Fig.\ref{fig:ideal_case_F_g}.
However we expect that in more realistic situations, e.g., when thermal fluctuations are present, this will not necessarily be the case. 
This will be investigated in Sec.~\ref{sec:limits_A} and Sec.~\ref{sec:limits_B} where we show the effect of non-zero vibrational excitation in the initial state, $\bar{n}>0$, when implementing $M_{r,j}^{A}$ and $M_{r,j}^{B}$, respectively.

Before turning to such discussion of imperfections, we continue here with the ideal case, considering the special case of equal amplitude-decay probabilities on both ions, $p_{1}=p_{2}=p$. In this case we can find an analytical expression for the unfiltered average gate fidelity independently of the scheme, which is given by
\begin{equation}
    \begin{split}\label{eq:F_avg_unf}
        &\bar{\mathcal{F}}_{g}^{\rm unf}=\frac{1}{5}\Big(1+ \frac{(1+\sqrt{\bar{p}})^{4}}{4}\Big),
    \end{split}
\end{equation}
and clearly depends only on the error probability. The analogous quantity for the filtered case with identical strength of the reversal operations on both ions set equal to the probability of decay, i.e., $p_{r}=p$ is  given by
\begin{equation}
\begin{split}\label{eq:F_avg_f}
\bar{\mathcal{F}}_{g}^{\rm f}=&\frac{1}{5}\Big(1+\frac{16\bar{p}^{2}}{4P_{r}}\Big)=\frac{1}{5}\Big(1+\frac{16}{(2+p)^{2}}\Big)
\end{split}
\end{equation}

Here the expression for the success probability that appears in the denominator is $P_{r}=\bar{p}^{2}(2+p)^{2}/4$.

Fig.~\ref{fig:ideal_case_F_g} shows plots of the analytic expressions for gate fidelity in the ideal case, together with the corresponding numerical simulations as a function of the gate 
time, where this is measured in units of $t/T_{1}$. 
The dashed black line represents the analytic unfiltered average gate fidelity (Eq. \ref{eq:F_avg_unf}), while the solid black line is the analytical average gate fidelity when the filtering is applied.
The fidelity under filtering always lies above the unfiltered case, and thus shows the benefits of the weak measurement reversal.  The dot and cross markers represent the results for the numerical simulations according to Scheme A (QSL-based) and Scheme B (entangling based), respectively.  

As mentioned in Sec.~\ref{sec:introduction}, where we discussed an example of quasi-distillation into the Bell pair  $\ket{\Psi_-}$, the advantage of the proposed scheme is more significant at longer times. It is interesting to note that in the limit of large $t$, the unfiltered average gate fidelity drops below $1/2$ and reaches the limit of $\bar{\mathcal{F}}_{g}^{\rm unf}=0.25$, whilst the filtered gate fidelity never drops below the value $\bar{\mathcal{F}}_{g}^{\rm f}= 0.55$.
Note that, for any biseparable state, the fidelity $F\leq 1/2$ whereas the fidelity is always greater than one-half, $F>1/2$, for entangled states \cite{Terhal2000, RevModPhys.81.865}. Thus, by applying the filtering operation, we can ensure that the system qubits remain in an entangled state. This can be easily characterized in current experiments using stabilizer-based witnesses \cite{Toth2005,PRXQuantum.2.020304, PhysRevX.12.011032} without the need to measure the full density matrix \cite{nielsen00}. Despite these positive features however, the probability does always decrease with time, as depicted by the blue solid line in Fig.~\ref{fig:ideal_case_F_g}. So for large $t$, it is unlikely that one can reverse the amplitude decay and recover the maximally-entangled states.

\begin{figure}[ht!]
    \includegraphics[width=1\columnwidth]{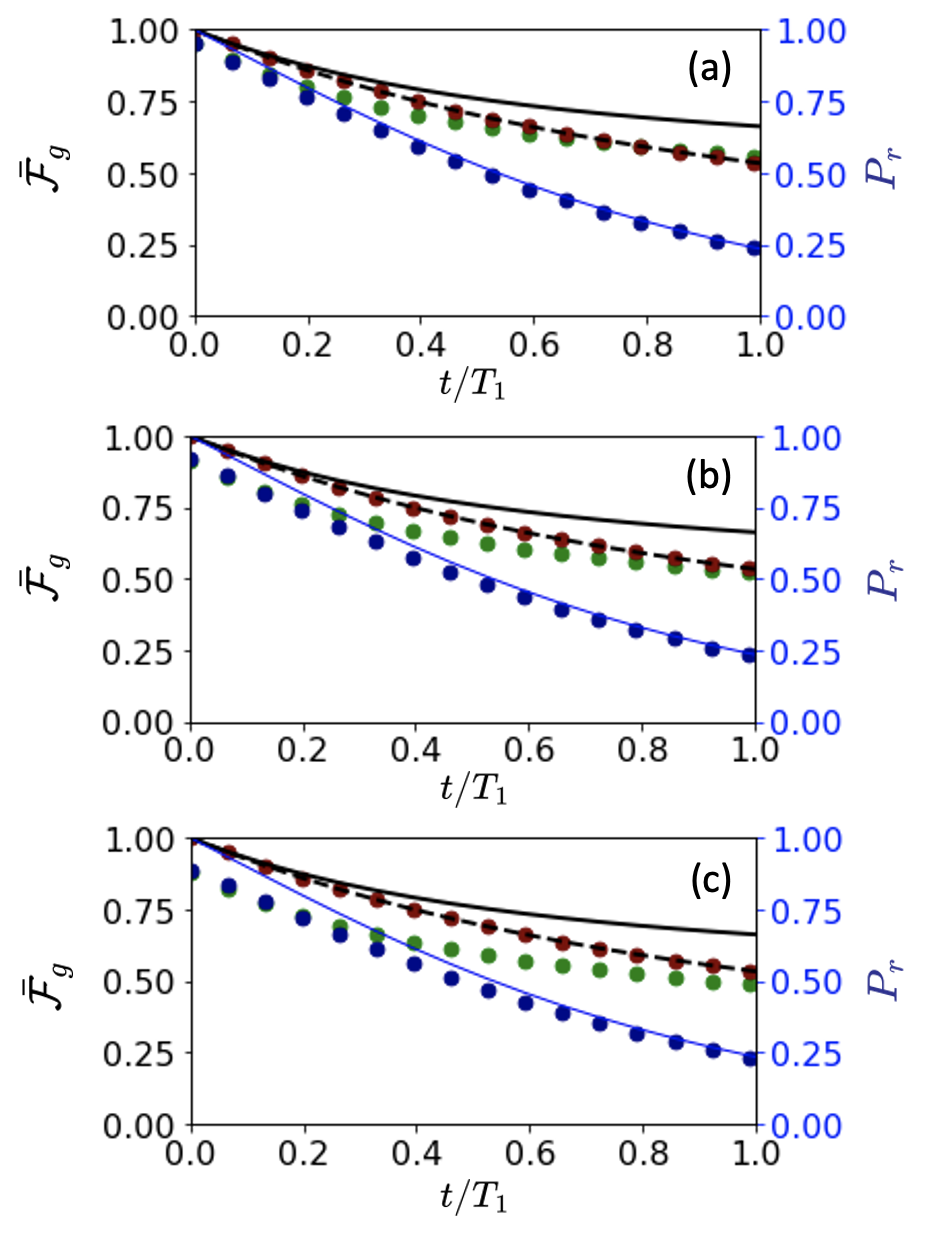}\\
    \caption{{\bf Average gate fidelities and success probability for scheme A with $\bar{n}>0$:} Numerical simulations of scheme A for two motional modes with average phonon number $\bar{n}\s{m}=\bar{n}\s{m_1}=\bar{n}\s{m_2}>0$. 
    Instead of starting with two Fock states with $\bar{n}=0$ as in the ideal case, we initialize the motional modes using two Gibbs states (\ref{eq:Gibbs}) with three different values of finite average phonon number: ($\rm{a}$) $\bar{n}\s{m}=0.05$, ($\rm{b}$) $\bar{n}\s{m}=0.09$ and ($\rm{c}$) $\bar{n}\s{m}=0.125$. 
    The black font y-axis on the left represents the average gate fidelity and the blue y-axis on the right represents the success probability $P_r$. The x-axis represents time in units of the amplitude damping time $T_{1}$. $T\s{1}=0.8$ s for the results shown here. The black solid lines represent the analytic and ideal-case average gate fidelities 
    when the filtering protocol is implemented after the unitary, i.e., $\bar{\mathcal{F}}_{g}^{\rm f}$. The dashed black lines are the analytical and ideal-case bare average gate fidelities, and the blue solid lines are the result of the success probabilities of the process simulated for different occupation numbers. The scattered green, maroon and blue dots represent the numerically simulated results under non-ideal conditions for the filtered average gate fidelity $\bar{\mathcal{F}}_{g}^{\rm f}$, the unfiltered average gate fidelity $\bar{\mathcal{F}}_{g}^{\rm unf}$, and the success probability $P_r$, respectively.}
    \label{fig:schemeA_limits} 
\end{figure}

\subsection{Limitation on Scheme A: warm vibrational modes }\label{sec:limits_A}

The QLS-based protocol, scheme A, was discussed above in the limit of zero  occupation number for both vibrational modes. However, in current experimental architectures, one commonly has $\bar{n}>0$ and the motional modes should then be described by the tensor product $\rho\s{\{m_q\}}^{\rm th}=\rho\s{m_{1}}^{\rm th}\otimes\rho\s{m_{2}}^{\rm th}$ of two Gibbs states of the form
\beq\label{eq:Gibbs}
  \rho\s{m_{q}}^{\rm th}=\sum_{n\s{m_q}=0}^{\infty}p\s{m_q}(n\s{m_q})\ket{n\s{m_q}}\bra{n\s{m_q}},
\eeq
where the probability $p\s{m_{q}}(n\s{m_q})$ is given by the thermal mode distribution
\begin{equation}
   p\s{m_{q}}(n\s{m_q})=\frac{1}{1+\bar{n}\s{m_q}}\left(\frac{\bar{n}\s{m_q}}{1+\bar{n}\s{m_q}}\right)^{n\s{m_q}}.
\end{equation}
Here $\bar{n}\s{m_q}=1/(\ee^{{k_{\rm B}T\s{m_q}/\hbar \omega\s{m_q}}}-1)$ is the Bose-Einstein distribution,  $\omega\s{m_q}$ is the frequency of the $m\s{q}$ mode, and $T\s{m_q}$ is an effective temperature for the mode. The resulting thermal fluctuations in the initial state will introduce errors in the system-ancilla mapping, and hence in the subsequent 
post-selected measurements on the ancillas, leading to imperfect filtering. To characterise these deviations, we numerically simulate the same sequence of unitaries discussed for the ideal case, with the vacuum motional state replaced by the thermal state~\eqref{eq:Gibbs}.

Fig.~\ref{fig:schemeA_limits} shows how both the fidelity after the probabilistic filtering, $\bar{\mathcal{F}}_{g}^{\rm f}$, and the success probability $P_r$ decrease as the mean number of vibrational excitations  $\bar{n}$ grows.  
Panels ($\rm{a}$), ($\rm{b}$) and ($\rm{c}$), show the results obtained assuming that the two vibrational modes exploited for QLS of each system-ancilla ions have been sideband-cooled to the same temperature, with different mean numbers of phonons $\bar{n}=0.05$, $\bar{n}=0.09$, and $\bar{n}=0.125$, respectively. Comparing with the results for the ideal case $\bar{n}=0$ (black-solid lines), we see how the filtered average gate fidelity $\bar{\mathcal{F}}_{g}^{\rm f}$ (scattered green dots) degrades significantly for even a small increase in the initial vibrational excitation number $\bar{n}0$. It is interesting to note that, when the amplitude decay is not too large, with this imperferction the gate fidelity after noise filtering (scattered green dots) can actually be worse than for the unfiltered case (dashed black-lines and scattered maroon dots). This is visible already for $\bar{n}>0.05$, where the filtering method is seen to provide no advantage for times $t/T\s{1}<1/2$ - here the unfiltered average gate fidelity (green dots) lies above  the filtered one (dashed black-lines and scattered maroon dots). This effect becomes larger as the mean number of phonons $\bar{n}$ increases.  For $\bar{n}>0.125$, the filtering barely adds any advantage, and it even drops below $1/2$ when $t\rightarrow T\s{1}$. The success probabilities (scattered blue dots) also drop with increasing average phonon number, such that one would get a reduced number of post-selected events. These results clearly show how sensitive the QLS-based scheme A is to thermal motion. For this reason it is imperative to have an alternative method such as the entanglement-based scheme B.  In the next subsection we discuss the performance of scheme B in the presence of its most serious source of imperfections.
\subsection{Limitation on entanglement based scheme B: warm active and spectator modes along the trap axis}\label{sec:limits_B}
As already mentioned previously, scheme B does not rely on motional Fock states having a single motional excitation like scheme A. Instead, scheme B exploits the two-qubit gates $U^{ZZ}(-\chi_{\rm c})=\ee^{\ii\chi_{\rm c}\sigma^z\s{a_q}\sigma^z\s{s_q}}$ between system and ancilla qubits, that
is significantly less sensitive to the thermal populations. The $U^{ZZ}(-\chi_{\rm c})$ gates implemented in the scheme are robust with respect  to thermal occupation of the active vibrational modes in the ideal scenario described in Sec. \ref{sec:schemeB}. This ideal scenario is the one in which the laser beams are perfectly aligned along one out of the three branches of phonons (one branch per trap symmetry axis $\alpha=x,y,z$). With $N$ 
ions, we have $N$ normal modes of oscillation per branch, each with an associated normal mode frequency $\omega_{\alpha, m\s{q}}$ and $m\s{q} \in {1,...N}$. We choose a particular axis of vibrational motion $\alpha$, and mode $m\s{q}$ to implement the entangling gates. 
This is the active mode, while the rest of the $(N-1)$ modes on that axis are considered as spectator modes. 
In the ideal scenario, it is assumed that the residual qubit-phonon

coupling with the modes that do not participate in the state-dependent force can be neglected. 
There are, however, corrections of a higher order in the Lamb-Dicke parameter, $\eta\s{m\s{q}}$, which are not far off-resonant and must be considered as a possible source of errors. Here, we study how such coupling of the warm active and spectator modes along the trap axis to the active entangling mode affect the performance of scheme B. 
The geometric phase $\chi_{c}$ of the entangling gates between ancilla and system qubits relates to the spin-spin coupling strength $J\s{a_q,s_q}$ and the gate time $t_{g}$ according to $J\s{a_q,s_q}t_{g}=-\chi_{c}$, with \cite{Milburn2000}
\beq
U\s{a_q,s_q}(t_{g})=\ee^{-\ii J\s{a_q,s_q}t_{g}\sigma^z\s{a_q}\sigma^z\s{s_q}}=\ee^{\ii\chi_{\rm c}\sigma^z\s{a_q}\sigma^z\s{s_q}}.
\eeq
where $\chi_{c}=\cos^{-1}(\sqrt{1-p_{r}})$. 
In the presence of warm modes, the coupling strength takes the form 
\begin{equation}
\label{eq:error_schemeB}
\begin{split}
   & J\s{a_q,s_q}=-\sum_{m_q}^{N}\frac{2}{\delta\s{m_q}}\frac{|\tilde{\Omega}\s{L,a_q}||\tilde{\Omega}\s{L,s_q}|}{4}\eta\s{m_q}^{2}M\s{a_q,m_q}M\s{s_q,m_q}\\
    &\times\left(1-\eta\s{m_q}^{2}a\s{m_q}^{\dagger}a\s{m_q}-\sum_{m'_q\neq m_q}\eta\s{m'_q}^{2}a\s{m'_q}^{\dagger}a\s{m'_q}\right)\mathrm{cos}(\phi\s{a_q,s_q}^{\circ}),
\end{split}
\end{equation}
where $\tilde{\Omega}\s{L,i}=\Omega\s{L}  \ee^{-\frac{1}{2}\sum\s{m_q}(\eta\s{m_q}\mathcal{M}\s{i,m_q})^{2}}$ includes the crossed-beam ac-Stark shifts   of  the transition  $\Omega\s{L}$, multiplied by the Debye-Waller factor \cite{PhysRevA.62.022311, PhysRevX.7.041061}, which accounts for a renormalization due to the zero-point fluctuations of the ions. Here  $M_{i,m_q}$ are the normal-mode displacements of ion $i$ along a given axis-direction for the $m\s{q}$-th mode \cite{PhysRevLett.110.110502}, $\delta\s{m_q}$ are the detunings of the laser frequency from the different mode frequencies, $a\s{m_q} (a'\s{m_q})$ is the annihilation operator for the active (spectator) modes and  $a^{\dagger}\s{m_q}(a'^{\dagger}\s{m_q})$ are the annihilation operators of the active (spectator) phonons in the common vibrational mode $n\s{m_q}$, and $\eta\s{m_q}$ is the Lamb-Dicke parameter. 

This analysis shows that thermal fluctuations on the vibrational modes will lead to deviations from the target condition $J\s{a_q,s_q}t_{g}=-\chi_{c}$. Eq.~(\ref{eq:error_schemeB}) predicts
the leading error due to thermal phonons on the active mode to be  $\eta\s{m_q}^{2}a\s{m_q}^{\dagger}a\s{m_q}$ which is $O(\{ \eta\s{m_q}^{2}\})$, and the leading error due 
to $N-1$ warm spectator modes along the same axis $\alpha$ to be 
$\sum_{m'_q\neq m_q}\eta\s{m'_q}^{2}a\s{m'_q}^{\dagger}a\s{m'_q}$ 
which is similarly $O(\{ \eta\s{m'_q}^{2}\})$.
These two corrections will add fluctuations to the phase of the entangling gate. However, the geometric phase closure conditions are not modified by thermal fluctuations. The form of the phase-space distribution may change, but the phase-space trajectory still closes  after the same time 
$t_{g}=2\pi r/\delta\s{m_q}$, where $r$ is the number of loops in phase space. If we fix the laser intensities such that, after this time, the area acquired in phase space leads to the desired $\chi_c$, we can estimate how scheme B gets affected by the average number of phonons using the relation
\begin{equation}
\label{eq:new_strength}
-J_{i,j}(\{ \bar{n}\s{m_q}\})t_{g}=\chi_{c}=\mathrm{cos}^{-1}(\sqrt{1-p_{r}})O(\{ \bar{n}\s{m_q}\}).
\end{equation}
where the leading corrections $O(\{\bar{n}\s{m_q}\})$ due to warm phonons on the trap axis direction are given by
\begin{equation}\label{eq:correction_schemeB}
\begin{split}
&O(\{\bar{n}\s{m_q}\})=1-\eta\s{m_1}^{2}\\
&\times\left[\bar{n}\s{m_1}+\left(\frac{\omega\s{m_1}}{\omega\s{m_2}}\right)\bar{n}\s{m_2}+\left(\frac{\omega\s{m_1}}{\omega\s{m_3}}\right)\bar{n}\s{m_3}+\left(\frac{\omega\s{m_1}}{\omega\s{m_4}}\right)\bar{n}\s{m_4}\right].
\end{split}
\end{equation}
 Here
 $\bar{n}\s{m_q}$ represent the average number of phonons on each mode.
 This expression is specific to the $N=4$ ion-chain depicted in Fig.~\ref{fig:general_filtering_second_scheme}. 
 For a given axis 
 $\alpha$, we choose the active mode to be the center-of-mass (COM) mode and set the frequency of this as $\omega\s{m_1}=\omega\s{COM}$.
 According to \cite{James1998}, the frequencies for the remaining (N-1) axial modes in an $N=4$ chain are $\omega\s{m_2}=\sqrt{3}\omega\s{m_1}$, $\omega\s{m_3}=\sqrt{5.81}\omega\s{m_1}$ and $\omega\s{m_4}=\sqrt{9.308}\omega\s{m_1}$. 

\begin{figure}[ht!]
    \includegraphics[width=1\columnwidth]{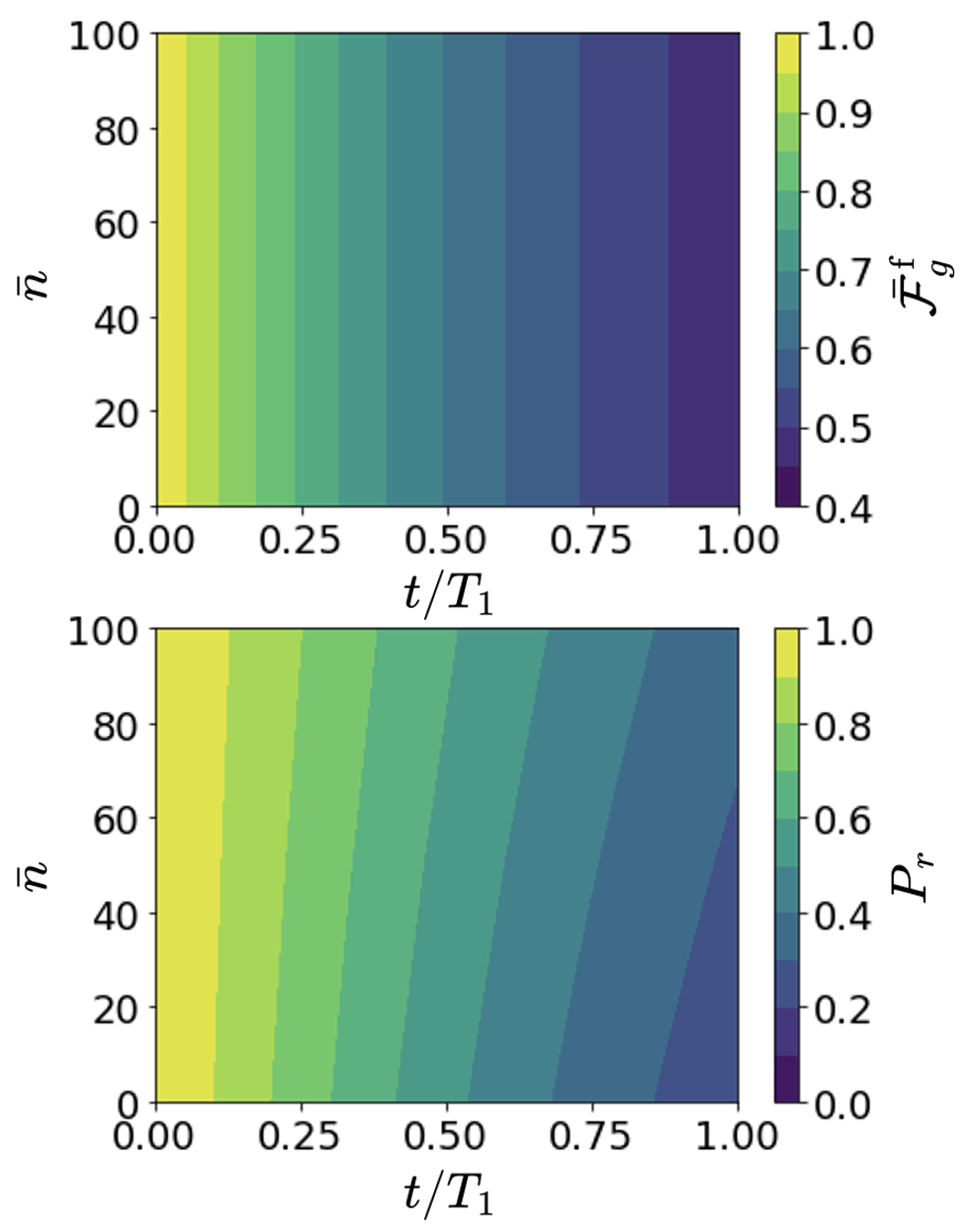}\\
    \caption{{\bf Effect of warm active phonons along the trap axis for scheme B:} 
    Numerical results for  simulation of scheme B following the sequence from Fig.(\ref{fig:general_filtering_second_scheme}) with  $\chi_{c}=\mathrm{cos}^{-1}(\sqrt{1-p_{r}})O(\{ \bar{n}\})$ as the strength of the measurement reversal, and $O(\{ \bar{n}\})$ (\ref{eq:correction_schemeB}) the leading corrections due to the warm phonons.
    The y-axes represent the average phonon number for each of the four modes which we set to be equal $\bar{n}=n\s{m_1}=n\s{m_2}=n\s{m_3}=n\s{m_4}$ whilst the x-axis represents the time in units of the amplitude damping time $T_{1}$. For the results shown in here we considered $T\s{1}=0.8$ s.
    The colormap sidebar on the top pannel represent the filtered/distilled average gate fidelity for different $\bar{n}$ and $t/T\s{1}$ units of time. The colormap on the bottom pannel does the same but for the probability of success.}
    \label{fig:limit_schemeB} 
\end{figure}

Fig.~\ref{fig:limit_schemeB} shows the result of a numerical simulation of the reversal operation $M_{r}=M_{r}^{B}$ using the strength from Eq.~(\ref{eq:new_strength}) for a linear chain with $N$=4 ions of the same species confined in a Paul trap having an axial COM frequency of $\omega\s{COM}=2\pi\times 1.4$ MHz and a Lamb-Dicke parameter of $\eta_{1}\equiv\eta\s{COM}=0.026$. The figure presents a contour plot for the simulated filtered gate fidelity $\bar{\mathcal{F}}_{g}^{\rm f}$ as a function of the amplitude decay time $T_1$ and the average number of phonons $\bar{n}$, where we considered the four modes to have the same occupation phonon number $\bar{n}=n\s{m_1}=n\s{m_2}=n\s{m_3}=n\s{m_4}$.

We see that
for times $t/T\s{1}<1/2$, the effect of higher phonon numbers $\bar{n}$ barely affects  the filtered  gate fidelity. In this situation, the reversal operations following the implementation of scheme B are quite robust against thermal motion.  
On the other hand, for longer decay times $t\rightarrow T\s{1}$, the filtered fidelity worsens slightly as one increases the average phonon number $\bar{n}$. Nevertheless, this scheme is considerably  more robust than scheme A. It is also noteworthy to discuss the effect of increasing $\bar{n}$ on the success probability $P_{r}$. In this case, the chance of a correct reversal increases with larger values $\bar{n}$. The effect is most significant for larger $t/T\s{1}$, but is still appreciable for shorter times. This means that we can fix a time $t$, and the  higher temperatures may help to increase slightly the probability of success without affecting the average gate fidelity, thereby improving the efficiency of the method. 

We conclude that the presence of warm active and spectator phonons on the trap axis direction could limit the distillation capabilities of Scheme B when $t\rightarrow T\s{1}$ but it has almost no effect at shorter times, even for $\bar{n}>50$. High $\bar{n}$ can even become advantageous in terms of increasing the success probabilities when the probability of amplitude decay is small, i.e., $T_1$ is large.

\section{Conclusions and Outlook}
\label{sec:conclusion}

We have presented a method to perform probabilistic  suppression  of amplitude damping to protect any maximally-entangled pair of trapped ion qubits from spontaneous photon scattering taking place during or after two-qubit entangling gates. The proposed  method can be understood as a non-unitary filter that allows for single-copy quasi-distillation. It can be applied to situations where the physical qubits are distant and where performing the conventional stabiliser readout of  encoded quantum states is not straightforward.
We have shown that such a filter can be obtained by post-selecting on a weak measurement implemented via ancillary ions. The filter can be realized by either a QLS-based scheme or an entangling-gate scheme.  In both cases, the non-unitary filter helps to reduce the overhead in the number of physical and ancillary qubits that is typically found in even the smallest QEC codes \cite{PhysRevA.56.2567,4675715}. It can therefore be visualized as an alternative protection method that  complements other error suppression approaches such as dynamical decoupling or encoding into decoherence-free subspaces. The method is clearly useful for platforms that focus on a small number of trapped ions such as clocks, sensors or quantum repeaters. 
We have also analyzed the role of thermal fluctuations on the amplitude damping reversal, which generate a possible source of noise.  We showed that such fluctuations can constitute a limiting factor in the QLS scheme, while the entangling gate scheme is not only less sensitive to thermal fluctuations but may benefit from these at longer operation times. 
Other possible limitations to long storage times of trapped ions may 
stem from decay of the auxiliary metastable excited $\ket{r}$ state in the lambda configuration used in scheme A for the realization of the quantum measurement reversal POVM, 
and from the choice of  ancilla qubits. These issues can be overcome by choosing $\ket{r}$ states with larger relaxation times and using ancillas from a different isotope or atomic species. Another limitation may be imposed by the heating rate of the vibrational modes, especially for the QLS-based method that requires the qubits to remain in the ground-state of motion throughout the protocol. 
Future studies directed towards laboratory implementation of the protocol should consider these additional error sources, as well as the effects of control imperfections in the various pulses. Another interesting application of the present amplitude damping reversal achieved by the non-unitary filter could be the detection of qubit leakage.

\acknowledgements 
A.R.B acknowledges support by the Universidad Complutense de Madrid-Banco Santander Predoctoral Fellowship, the Fulbright Predoctoral Scholarship program (Fulbright Spain 2019-2020), and by GRADIANT, ICT R\&D centre in Galicia. A.R.B also thanks J.G.F.U for providing access to computing capabilities to carry out the numerical simulations.
A.B. acknowledges support from PID2021-127726NB-I00  (MCIU/AEI/FEDER, UE), from the Grant IFT Centro de Excelencia Severo Ochoa CEX2020-001007-S, funded by MCIN/AEI/10.13039/501100011033, from the grant QUITEMAD+ S2013/ICE-2801, and from the CSIC Research Platform on Quantum Technologies PTI-001.
K.B.W. was supported by the NSF QLCI program through grant number QMA-2016345.

\bibliography{weak_rm}

\end{document}